\documentclass[useAMS,usegraphicx,usenatbib]{mn2e}
	\usepackage[english]{babel}
	\usepackage[utf8]{inputenc}
	\usepackage[T1]{fontenc}
	\usepackage{amsmath,txfonts,graphics,graphicx,epstopdf,float,lscape,longtable,dcolumn,color,caption,subcaption,footnote,threeparttable,rotating,slashbox,multirow,url}
	
	\usepackage{natbib}
		\bibpunct{(}{)}{,}{a}{}{,}
	\pdfoutput=1

\def\arcsec {\hbox{$^{\prime\prime}$}}

\title[What is the most reliable tracer of core collapse in dense clusters?]{Line Profiles of Cores within Clusters. III. What is the most reliable tracer of core collapse in dense clusters?}
\author[R.-A.~Chira et al.]{Roxana-Adela~Chira$^{1,2,3,}$\thanks{E-mail: rox.chira@gmail.com}, Rowan~J.~Smith$^1$, Ralf~S.~Klessen$^1$, Amelia~M.~Stutz$^3$, \newauthor Rahul~Shetty$^1$ \\
	$^1$Universit\"at Heidelberg, Zentrum f\"ur Astronomie, Institut für Theoretische Astrophysik, Albert-Ueberle-Str. 2, 69120 Heidelberg, Germany\\
	$^2$European Southern Observatory, Karl-Schwarzschild-Str. 2, 85748, Garching b. München, Germany \\
	$^3$Max Planck Institute for Astronomy, K\"onigstuhl 17, D-69117 Heidelberg, Germany}
\date{Submitted 21.02.2014, Accepted 24.07.2014}
\pagerange{\pageref{firstpage}--\pageref{lastpage}} \pubyear{2014}

\begin{document}
\label{firstpage}
\maketitle
\begin{abstract}
	Recent observational and theoretical investigations have emphasised the importance of filamentary networks within molecular clouds as sites of star formation.
	Since such environments are more complex than those of isolated cores, it is essential to understand how the observed line profiles from collapsing cores with non-spherical geometry are affected by filaments.
	In this study, we investigate line profile asymmetries by performing radiative transfer calculations on hydrodynamic models of three collapsing cores that are embedded in filaments.
	We compare the results to those that are expected for isolated cores.
	We model the five lowest rotational transition line (J = 1-0, 2-1, 3-2, 4-3, and 5-4) of both optically thick (HCN, HCO$^+$) as well as optically thin (N$_2$H$^+$, H$^{13}$CO$^+$) molecules using constant abundance laws.
	We find that less than 50\% of simulated (1-0) transition lines show blue infall asymmetries due to obscuration by the surrounding filament.
	However, the fraction of collapsing cores that have a blue asymmetric emission line profile rises to 90\% when observed in the (4-3) transition.
	Since the densest gas towards the collapsing core can excite higher rotational states, upper level transitions are more likely to produce blue asymmetric emission profiles.
	We conclude that even in irregular, embedded cores one can trace infalling gas motions with blue asymmetric line profiles of optically thick lines by observing higher transitions.
	The best tracer of collapse motions of our sample is the (4-3) transition of HCN, but the (3-2) and (5-4) transitions of both HCN and HCO$^+$ are also good tracers.
\end{abstract}


\section{Introduction}\label{sec_intro}

	\begin{figure*}
		\centering
		\begin{minipage}{\textwidth}
			\centering
			\includegraphics[width=\textwidth]{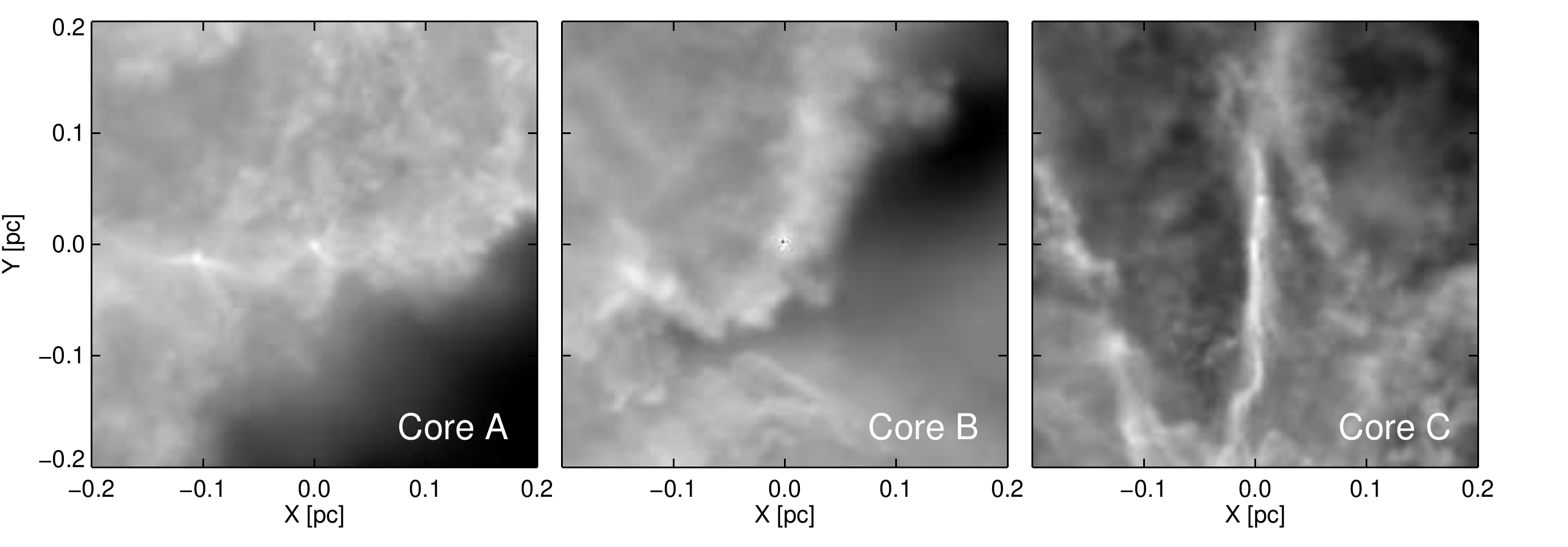}
			\caption{Core A, B and C in dust emission at 850 $\mu$m and at $i$ = 0$\degr$ and $\phi$ = 0$\degr$ \citep{Smith2012a}.}
			\label{fig_01_cores}
		\end{minipage}
	\end{figure*}

	While it has long been known that stars form in dense cores of gas within molecular clouds \citep[e.g][]{Myers1985,Cernicharo1991,Lada1993,Larson2003} it has only recently become appreciated that these cores are intrinsically connected to their environment through a network of filaments \citep[e.g][]{Menshchikov2010,Arzoumanian2011,Peretto2012}. 
	Our understanding of the physics and predicted observable properties of isolated cores is well developed, but the effects of surrounding filaments on observed core properties is still at an early phase of investigation. 
	This paper is the third in a series that attempts to investigate the predicted line profiles of collapsing cores embedded within filaments.

	Blue asymmetries in the line profiles of optically thick tracer molecules appear to be a good indicators of gas infall motions within isolated cores \citep{Zhou1991,Zhou1992,Walker1994,Myers1996}. 
	This effect arises because in optically thick species an observer sees the outer layers of the core on the red side of the line profile and the inner layers on the blue. 
	Using this effect several observers have tried to find nearby pre-stellar cores with infall motions. 
	\citet{Gregersen1997,Lee1999,Gregersen2000} and \linebreak \citet{Sohn2007} investigate samples of Class 0 and pre-stellar cores, which are potentially collapsing, and find that less than 65\% of the observed line profiles are blue asymmetric. 
	In general, blue line profiles are clearly detected in only about 40\% of the cores. 
	It is crucial to understand how many pre-stellar cores are collapsing in order to understand how rapid and dynamic the process of star formation is.
	However, it is not clear if a blue asymmetries should always be expected from a core embedded within a filamentary cluster of protostars. For a more detailed description of the theory behind the blue line asymmetry and the observational literature see, e.\ g., \citet{Evans1999} and \citet{Smith2012a}.

	In the first paper of our series about the line profiles of cores within clusters \citet{Smith2012a} (hereafter S12) investigate three filamentary cores, that are part of a dense giant molecular cloud (GMC) simulation, from 14 different viewing angles. 
	Though all the model cores are collapsing, less than 50\% of the \linebreak HCN (J = 1-0) profiles exhibited blue asymmetries.
	In \citet{Smith2013} this work is extended to look at massive star forming regions. 
	It is found that the line profiles frequently do not have a central self-absorption dip as the collapsing regions do not always have surrounding static envelopes. 
	Furthermore, in massive star forming regions multiple line components in optically thin species such as N$_2$H$^+$ are visible due to internal substructure. 
	The observation of such line profiles can be used to test models of massive star formation. 
	In this paper we seek to extend the analysis of S12 to consider additional line transitions and chemical species. 

	In S12 we consider the base (1-0) transitions for the sake of simplicity, since we already have a large number of variables. 
	However, one of the reasons the success rate of the blue asymmetry is so low in the study is that the filaments in which the cores are embedded are also brightly emitting in HCN (1-0) which obscures the emission from the core. 
	Species and transitions with a greater critical density than HCN (1-0) will only become bright at greater densities and temperature, and are consequently more likely to be visible in the core but not the surrounding filament. 
	It would be highly advantageous to an observer surveying dense cores to know in advance which species and line transitions are most likely to trace infalling gas in pre-stellar cores embedded within dense networks of filaments. 

	This is the question we attempt to answer in this paper: What is the most reliable tracer of core collapse in dense clusters? 
	In Section \ref{sec_methods} we describe the technical tools and set-ups which have been used for this study. 
	The results are presented and discussed in Section \ref{sec_results}. 
	Section \ref{sec_conclusion} provides a summary and final conclusions.


\section{Methods}\label{sec_methods}

\subsection{Molecular Line Modelling}\label{meth_radmc}
	We use the same three cores as S12 which are labelled as Core A, B and C and adopt this nomenclature here. 
	Fig.\ \ref{fig_01_cores} shows snapshot of their modelled dust emission at 850 $\mu$m. 
	The core centres are located in the middle of the snapshots (at $x$ = 0 pc and $y$ = 0 pc) where the dust emission peaks.
	Each of these three regions contains dense gas with number densities on the order \linebreak 10$^3$ to 10$^7$ cm$^{-3}$, total masses of 6.4 M$_\odot$, 12.3 M$_\odot$ and 14.7 M$_\odot$ and column densities on the order 10$^{22}$ to 10$^{23}$ cm$^{-2}$. 
	The gas temperature within the regions is on average 23 K.
	These rather high temperatures originate from the more diffuse gas but the dense gas has temperatures of around 10 K. 
	S12 confirmed that the actual temperatures do not change the qualitative behaviour of the line profile asymmetries by repeating their calculations using a constant gas temperatures of 14 K.
	All the properties of our sample cores are in good agreement with observed low-mass star-forming regions \citep{Bergin2007,Stutz2010,Hacar2011,Launhardt2013}. 

	S12 selected these three regions, because they contain locations of future sink-particles (representing the sites of future star formation).
	The cores are not spherically symmetric, but are collapsing and deeply embedded within networks of filaments. 
	The region around Core A represents a turbulent sheet, Core B forms at the interface between two regions with different velocities, and Core C is the result of a head-on colliding flow (S12).
	In S12 we concentrated on these three representative cores so that we could study the effects of projection upon our results.
	Since the distributions of gas density and temperature strongly depend on the actual line of sight angle we are able to cover a wide range of geometrical and dynamical features without the need for a higher number of target cores. 
	We purposefully limit our sample to collapsing cores so that the expected blue asymmetry rate should be 100\%.

	For line modelling we use the radiative transfer code 
	\texttt{RADMC-3D}\footnote{The code is available with the permission of the main author, Cornelis Dullemond, at the webpage \url{http://www.ita.uni-heidelberg.de/~dullemond/software/radmc-3d/}. 
	There is also a manual on the website including more detailed explanations of the different functions and parameters.}. 
	The code includes a treatment of scattering, thermal emission and absorption of both gas and dust. 
	Here we focus on modelling the line transfer of molecular tracers with high critical densities in dense molecular gas.
	We use the large velocity gradient (LVG) approximation (\citealt{Sobolev1957,Shetty2011a}) and the Doppler catching method, which resolves and smooths jumps in the velocity along the line of sight (\citealt{Shetty2011b}).
	\newpage

	We map the distributions of gas temperature, gas number density and large-scale velocity of the individual cores and their surrounding filaments to 200$^3$ grids which correspond to a physical volume of 0.4$^3$ pc$^3$.
	We assume a dust-to-gas ratio of 0.01 and a constant dust temperature of 20 K and set the level of microturbulence to 90 cm s$^{-1}$. 
	A test has proved that the actual value of the microturbulence does not influence the results significantly.

	We model the line profiles of each species and transition at 256 evenly-spaced wavelengths around the corresponding rest wavelengths. 
	For deriving the spectra we use a Gaussian beam with a FWHM of 0.01 pc which corresponds to about 13.7$\arcsec$ if the core is located at a distance of 150 pc.
	This matches the average diameter of the cores.
	We simulate the transition lines for each core and tracer molecule in twelve different line of sight-angles. 
	These correspond to eight sight-angles with inclinations $i$ between 0$\degr$ and 315$\degr$ in steps of 45$\degr$ at constant rotation angle $\phi$ = 0$\degr$ and four more with rotation angles $\phi$ between 45$\degr$ and 180$\degr$ in steps of 45$\degr$ at constant inclination $i$ = 90$\degr$.

	\begin{table*}
	\begin{minipage}{\textwidth}
	\centering
	\begin{threeparttable}
		\caption{A summary of properties of molecular tracers. We show the transitions, the adopted abundances (with references) and the corresponding critical densities for LTE; the latter being estimated by $n_{crit}\,=\,A_{ul} / K_{ul}$ where $A_{ul}$ represents the Einstein coefficient and $K_{ul}$ the collisional rate coefficient at a temperature of 20 K.}
		\label{tab01_abundance}
		\begin{tabular}{cc|c|c|cc}
		\hline
		Tracer & Transition Lines & Optically & Critical Density $n_{crit}$ & Abundance $\chi = n / n_{H_2}$ & Reference  \\ \hline
		& $J$ = &  & cm$^{-3}$ & & \\ \hline \hline
		N$_2$H$^+$ & (1-0)\tnote{a}, (3-2) & thin & 1.6 $\times$ 10$^5$, 3.0 $\times$ 10$^6$ & 1.0 $\times$ 10$^{-10}$ & \citet{Aikawa2005} \\
		H$^{13}$CO$^+$ & (3-2), (4-3) & thin & 1.7 $\times$ 10$^5$, 3.0  $\times$ 10$^6$ & 1.72 $\times$ 10$^{-11}$ & \citet{Maruta2010} \\ \hline
		HCN & (1-0) -- (5-4)\tnote{b} & thick & 1.0 $\times$ 10$^6$ -- 9.7 $\times$ 10$^8$ & 3.0 $\times$ 10$^{-9}$ & \citet{Lee2004} \\
		HCO$^+$ & (1-0) -- (5-4)\tnote{b} & thick & 1.8 $\times$ 10$^6$ -- 1.8 $\times$ 10$^7$ & 5.0 $\times$ 10$^{-9}$ & \citet{Aikawa2005} \\ \hline
		\end{tabular}
		\begin{tablenotes}
		\item[a] only using the isolated 101-012 hyperfine structure line
		\item[b] meaning $J$ = (1-0), (2-1), (3-2), (4-3), and (5-4)
	\end{tablenotes}
	\end{threeparttable}
	\end{minipage}
	\end{table*}

\subsubsection{Molecular Tracer Species}\label{meth_tracer}
	In our study we model the (1-0), (2-1), (3-2), (4-3) and (5-4) emission lines of both optically thick (HCN, HCO$^+$) and optically thin (N$_2$H$^+$, H$^{13}$CO$^+$) species. 
	Table \ref{tab01_abundance} summaries the transitions which we use for the discussion in this paper.
	It also provides the corresponding LTE critical densities.
	We take the line data that we need for computing the level populations from the LAMDA database \citep{Schoeier2005}.
	To simplify our analysis of the higher transitions we only use the data without hyperfine structure and molecular hydrogen as dominant collision partner for all species and transitions.
	We assume constant abundances for all species. 
	This is reasonable for the majority of our tracers, although it is a simplification for the densest parts deep within the cores \citep[e.\ g.,][]{Lippok2013}.
	The constant abundances allow us to focus on the effects of the core dynamics and density distribution on the line profiles. 
	We discuss how depletion may change this in Sect. \ref{res_abun}.
	
\subsection{Normalised Velocity Difference $\delta v$}\label{meth_deltav}
	In order to investigate line profile asymmetries, it is necessary to find a method with which spectra can be classified.
	One possibility is to classify the spectra individually by eye as blue, red asymmetric or ambiguous.
	But it is obvious that this method is not efficient for a large sample and may be a matter of taste in difficult cases.

	A more quantitative method is provided by the normalised velocity difference $\delta v$ method \citep{Mardones1997}. It is given by
	\begin{equation}
		\delta v\,=\,\dfrac{v_{thick} - v_{thin}}{\Delta v_{thin}} \label{equ_def_deltav}
	\end{equation}
	where $v_{thick}$ represents the central velocity of the higher peak in the line profile of the optically thick species, $v_{thin}$ the central velocity of the peak of the optically thin species (assuming the line profile has an Gaussian shape) and $\Delta v_{thin}$ its velocity dispersion. 
	Dividing the velocity difference with the velocity dispersion of the optically thin line normalises the difference and reduces the bias due to different line widths \citep{Mardones1997}.
	We obtain the values by fitting a single Gaussian to the optically thin lines and a linear combination of two Gaussians to the line profiles of the optically thick species.
	In our simulation $v_{thin}$ should be approximately 0 km s$^{-1}$ and is theoretically not needed. 
	However, we kept $v_{thin}$ as the reference, because observed velocities are composed of different components, like local standard of rest velocities etc.

	The higher peak central velocity of the optically thick species, $v_{thick}$ should differ from 0 km s$^{-1}$, either to positive or negative values depending on the asymmetry of the spectrum. 
	If the line profile has a red asymmetry the red peak, with positive central velocity, is higher. 
	Thus, $v_{thick}$ and $\delta v$ are positive. 
	On the other hand $v_{thick}$ and $\delta v$ become negative if the blue peak is higher and the line profile is blue asymmetric.
	If both peaks are equally bright, $\delta v$ is set to 0.

	The disadvantage of this method is that spectra with ambiguous symmetries, meaning line profiles which cannot clearly be identified as blue or red asymmetric (because the peaks are almost equally high), are hard to identify, because it detects even the smallest difference between the brightness temperatures. 
	To identify such cases we introduce two criteria, namely
	\begin{align}
		\dfrac{|T_{b,blue} - T_{b,red}|}{\max\{T_{b,blue},T_{b,red}\}} < 10\% \label{def_amb_cr1a} \\ \vspace*{0.5\baselineskip}
		| v_{thick,blue} - v_{thick,red} | < \Delta v_{thin} \label{def_amb_cr2a}
	\end{align}\label{def_amb}
	with $T_{b,i}$ being the peak brightness temperatures and $v_{thick,i}$ being the central velocity of the $i$th line component.
	The first criterion in equation (\ref{def_amb_cr1a}) represents line profiles where either an observer could not classify the profile clearly, or where observational difficulties, like noise, might cause problems (see Sect.\ \ref{res_obs}).
	The second criterion in equation (\ref{def_amb_cr2a}) makes sure that the double-peaked structure, if there is any, is resolved.
	For doing so, the separation between the peaks is compared to the FWHM of the reference line, $\Delta v_{thin}$. 
	If the separation is smaller than the FWHM it is not possible to guarantee that the separation is revolved accurately. 
	If either criterion is fulfilled, $\delta v$ is set to 0 and the spectrum is classified as ambiguous.

\subsection{Optical Depth Surfaces}\label{meth_tausurf}
	In our investigation we aim to study the origins of line profile features. 
	The simulated GMC data provide the distributions of gas density, velocity and temperature of our sample cores.
	As long as the observed gas is optically thin, we know that the emission originates from all the gas along the line of sight.
	But in the case of optically thick tracers, which are affected by self-absorption, the observed emission only originates from the closest point along the line of sight toward the observer at a given velocity component.
	To pinpoint these closest points we use the \texttt{RADMC-3D} function \texttt{tausurf}.
	This function returns the position where the optical depth, $\tau$, of the gas reaches a certain value for the first time along the line of sight at a given wavelength. 
	Here we look at the $\tau\,=\,1$ surfaces at the rest wavelengths of each transition. 
	
	We call the plots generated with the results of \texttt{tausurf} \textit{Optical Depth Surfaces} (ODS; see Figs.\ \ref{fig_08_multi_hcn} and \ref{fig_10_multi_n2h}) hereafter.
	They include two pieces of information. 
	The grey-scale background shows the emission of the tracer gas in logarithmic scales. 
	It acts as a reference for the general structure of the core and surrounding filaments as they would be seen by the observer. 
	The green star in the middle indicates the location of the core centre. 
	The coloured points in the foreground display the results of \texttt{tausurf}, thus the position where the tracer becomes optically thick. 
	The colours of the dots refer to the corresponding line of sight radius $z$, which connects the core at $z$ = 0 pc with the observer at $z\,\rightarrow\,-\infty$. 
	Red colours symbolise positions in front of the core and blue colours positions behind the core. 
	Thus, the redder the colour of a dot, the closer the surface is to the observer. 

	The advantage of the ODS is that they show which part of the total emitted light the observer actually observes. 
	With this information it is possible to set limits to the visible regions and to explain why features do not appear in the line profile although we had predicted them from looking at the density and velocity distributions only.

	\begin{figure*}
		\centering
		\begin{minipage}{\textwidth}
			\centering
			\includegraphics[width=\textwidth]{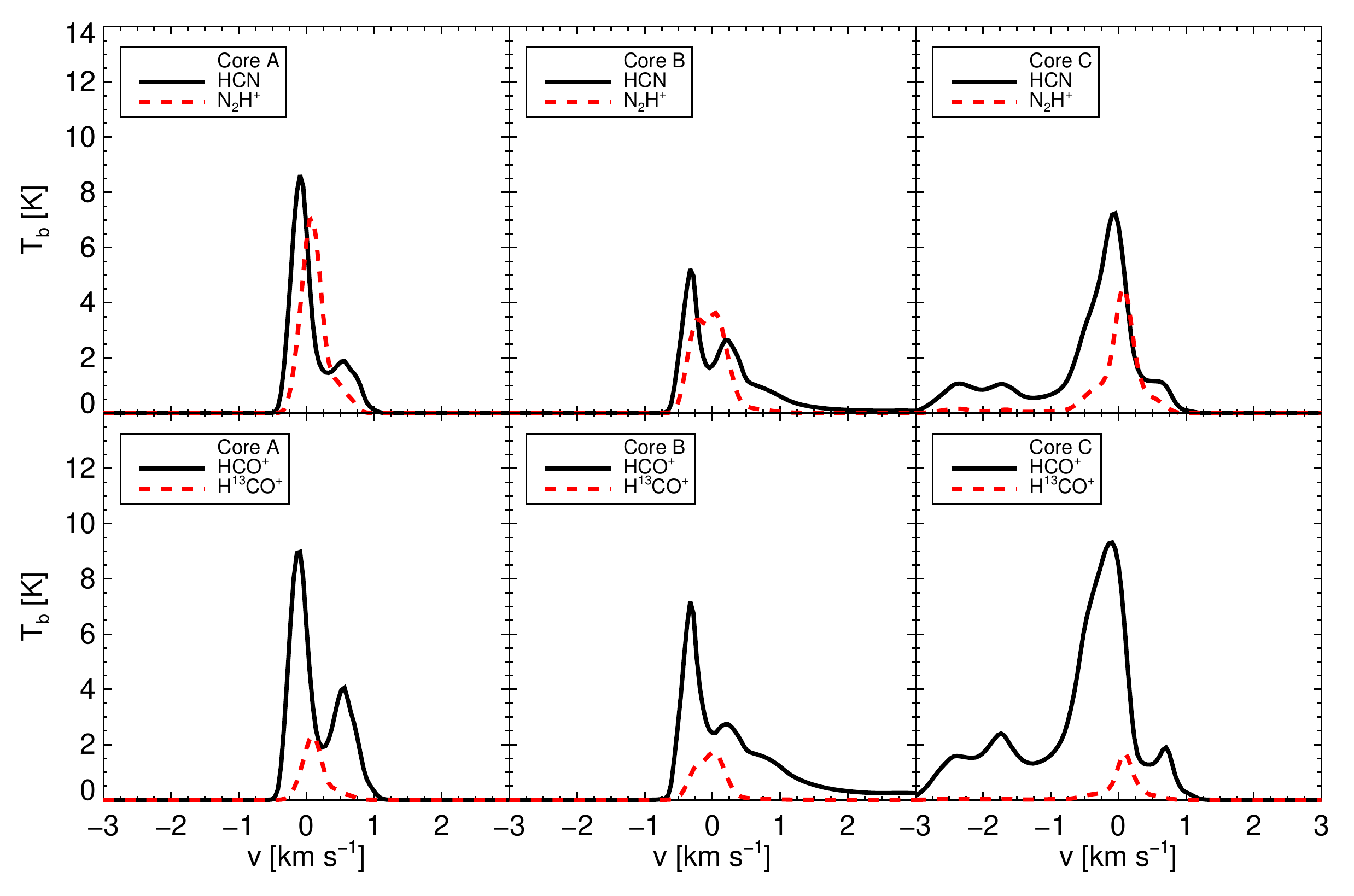}
			\caption{(1-0) transition lines of all tracer species in all cores at $i$ = 0$\degr$ and $\phi$ = 0$\degr$.}
			\label{fig_02_multi_all_10}
		\end{minipage}
	\end{figure*}
	
	\begin{figure}
		\vspace*{-\baselineskip}
		\centering
		\includegraphics[width=0.48\textwidth]{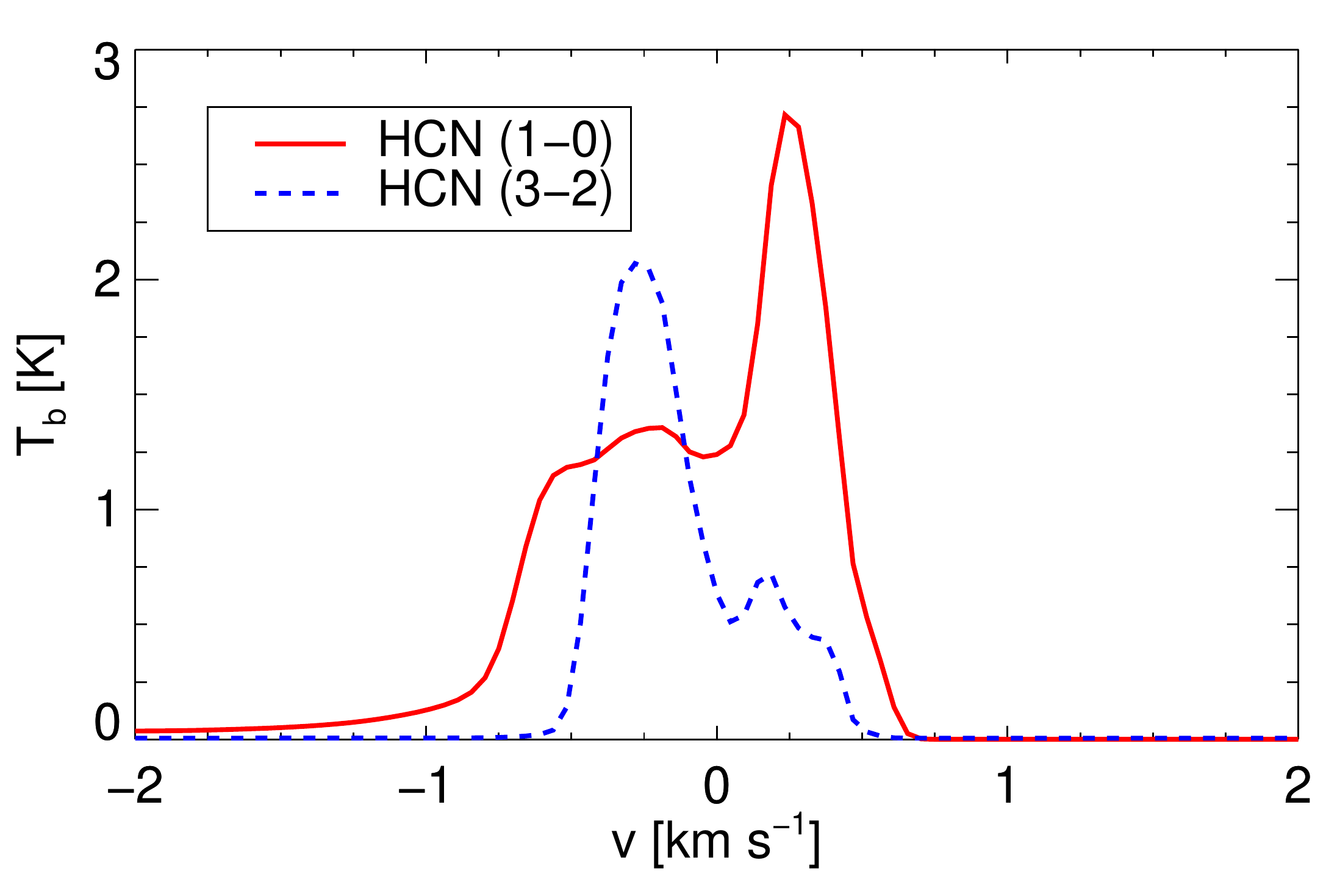}
		\caption{Example of observed line profiles of HCN (1-0) and (3-2). They are observed in Core B at $i$ = 90$\degr$ and $\phi$ = 135$\degr$. Both lines show different asymmetries. The line profile of HCN (1-0) is red asymmetric, whereas the line profile of HCN (3-2) is blue asymmetric.}
		\label{fig_03_red10_blue32}
		\vspace*{-\baselineskip}
	\end{figure}

\section{Results}\label{sec_results}

\subsection{Line Profiles}\label{res_profiles}
	Fig.\ \ref{fig_02_multi_all_10} shows the line profiles of the (1-0) transition of all species and cores seen at inclination and rotation angles of 0$\degr$. 
	The line profiles of HCN and HCO$^+$ show optical depth effects, such as the characteristic self-absorption feature.
	However, contrary to the theoretical model \citep{Evans1999} many spectra are not blue, but red asymmetric or ambiguous (as discussed by S12).
	The line profile asymmetries at given inclinations and rotation angles are not invariant.

	\noindent Comparing the line profiles at different transitions there are three most commonly occurring cases:
	\begin{description}
		\item[(i)] \hspace*{0.35em} all lines are blue asymmetric, which is the ideal case,
		\item[(ii)] \hspace*{0.1em} all lines are red asymmetric, which is the worst case, and
		\item[(iii)] the lines at lower transitions, (1-0) and (2-1), are red \linebreak \hspace*{2em} asymmetric, but the (3-2) and higher transitions are blue \linebreak \hspace*{2.25em} asymmetric (see Fig.\ \ref{fig_03_red10_blue32}), which is the most frequent case.
	\end{description}
	We observe such a behaviour for all optically thick species.
	The third case is the most interesting one, because it is directly connected to the question whether blue infall asymmetries are supposed to be observed in irregular cores.
	The first two cases alone would have negate the assumption of a particular asymmetry.
	The changes of asymmetries in line profiles imply that we can expect that blue asymmetric line profiles trace the infall motion of early star-forming cores.
	However, there are also mechanisms close the infalling material that may influence the observed line profile asymmetries such that they do not necessarily trace the gas motions within the core anymore.
	Consequently, conclusions about collapse motions within the core region may turn out to be erroneous when only assuming the simple case model of a spherically collapsing core.
	\smallskip

	In the case of optically thin tracers, the majority of N$_2$H$^+$ and H$^{13}$CO$^+$ line profiles appear Gaussian.
	But as Fig.\ \ref{fig_02_multi_all_10} shows there are exceptions where they are also affected by self-absorption. 
	Such non-Gaussianities render it difficult to compute the true rest velocity and introduce additional uncertainties to the parameter determination.
	If one cannot calculate the central velocities reliably, this has a significant impact on the results of the normalised velocity differences $\delta v$, see equation (\ref{equ_def_deltav}). 
	\smallskip
	
	Many line profiles, of both optically thick and optically thin tracers, have additional emission features which cannot be explained solely by the influence of optical thickness.
	Fig.\ \ref{fig_02_multi_all_10} shows examples of line profiles from Core C that consist of two components.
	There is a bright component around 0 km s$^{-1}$ which belongs to the central regions of the core. 
	The second component at about $-2$ km s$^{-1}$ originates from a second region of dense gas along the line of sight.

	\begin{figure*}
		\centering
		\begin{subfigure}{0.48\textwidth}
			\includegraphics[width=\textwidth]{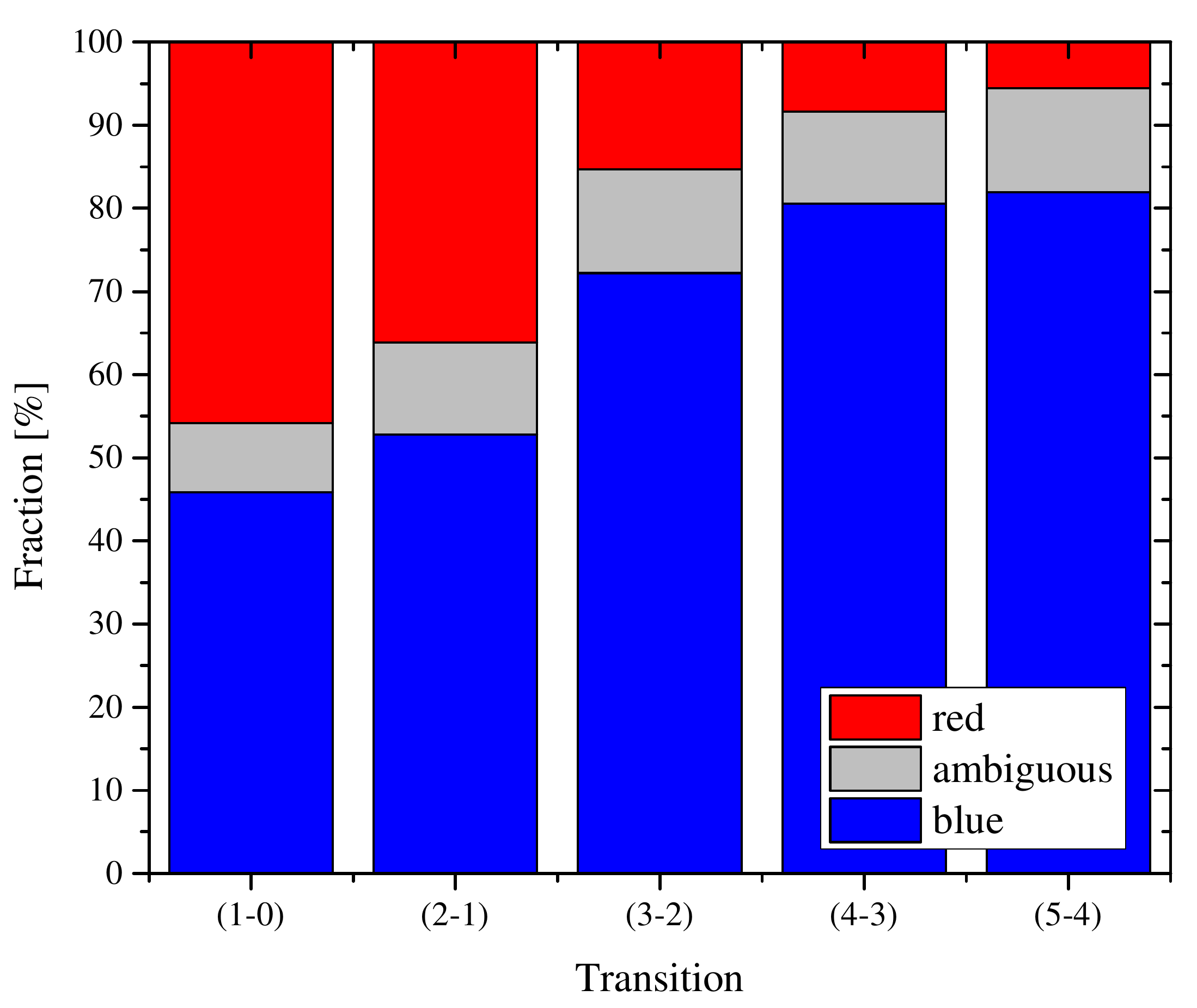}
			\caption{Results using the N$_2$H$^+$ (1-0) transition as velocity reference.}
			\label{fig_04_deltav_n2h10}
		\end{subfigure} \hfil
		\begin{subfigure}{0.48\textwidth}
			\centering
			\includegraphics[width=\textwidth]{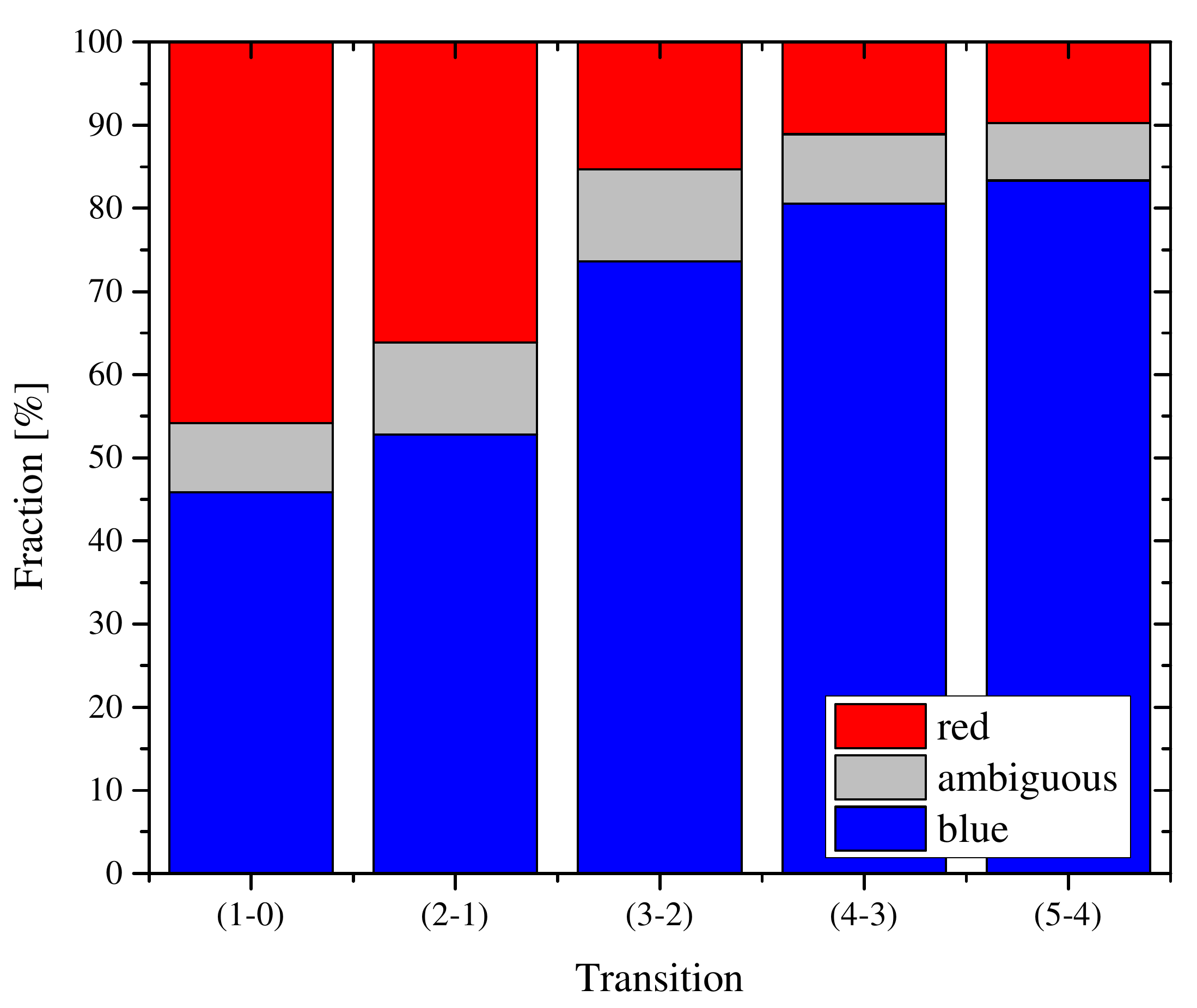}
			\caption{Results using the N$_2$H$^+$ (3-2) transition as velocity reference.}
			\label{fig_04_deltav_n2h32}
		\end{subfigure} \\
		\begin{subfigure}{0.48\textwidth}
			\centering
			\includegraphics[width=\textwidth]{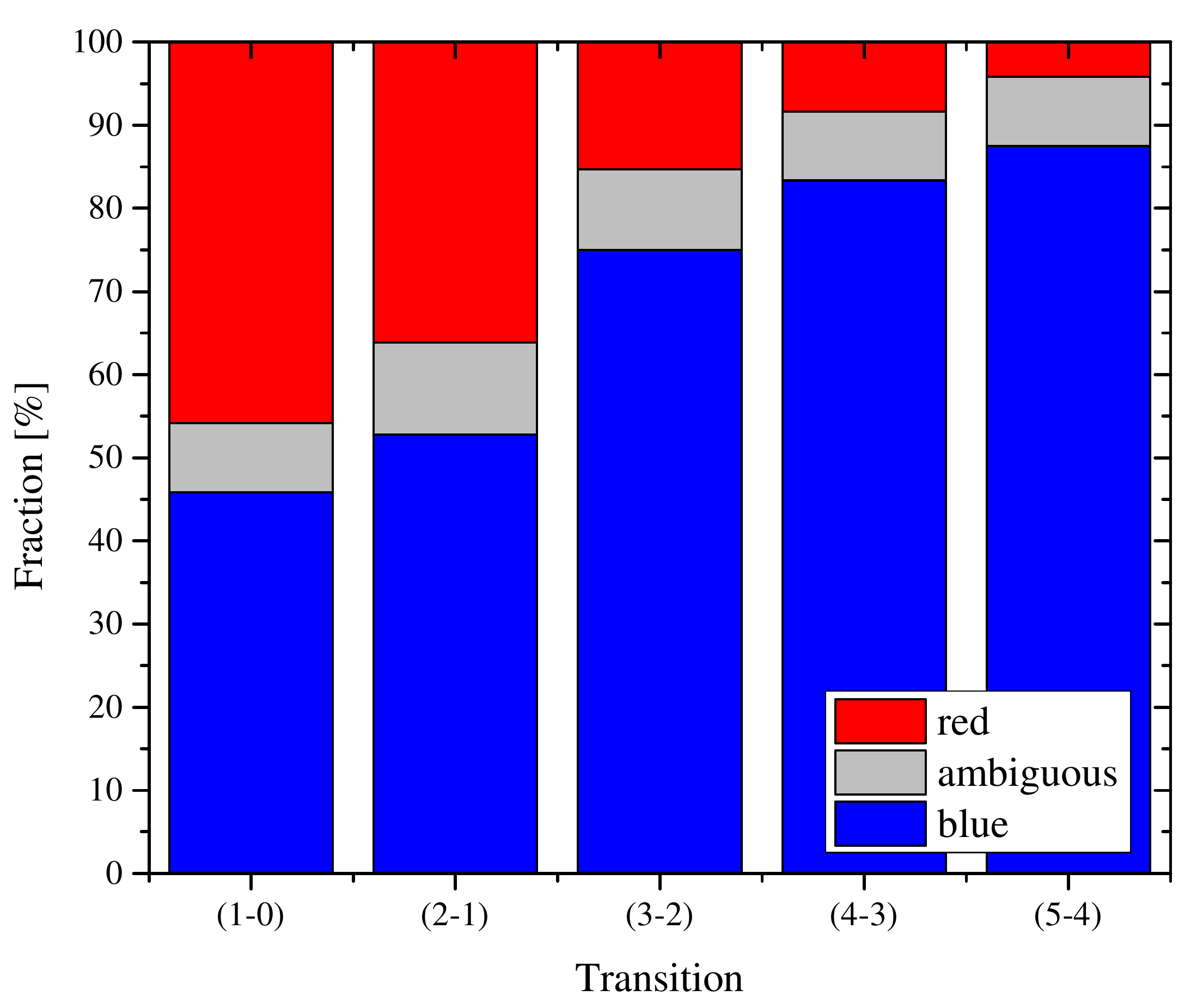}
			\caption{Results using the H$^{13}$CO$^+$ (3-2) transition as velocity reference.}
			\label{fig_04_deltav_h13co32}
		\end{subfigure} \hfil
		\begin{subfigure}{0.48\textwidth}
			\centering
			\includegraphics[width=\textwidth]{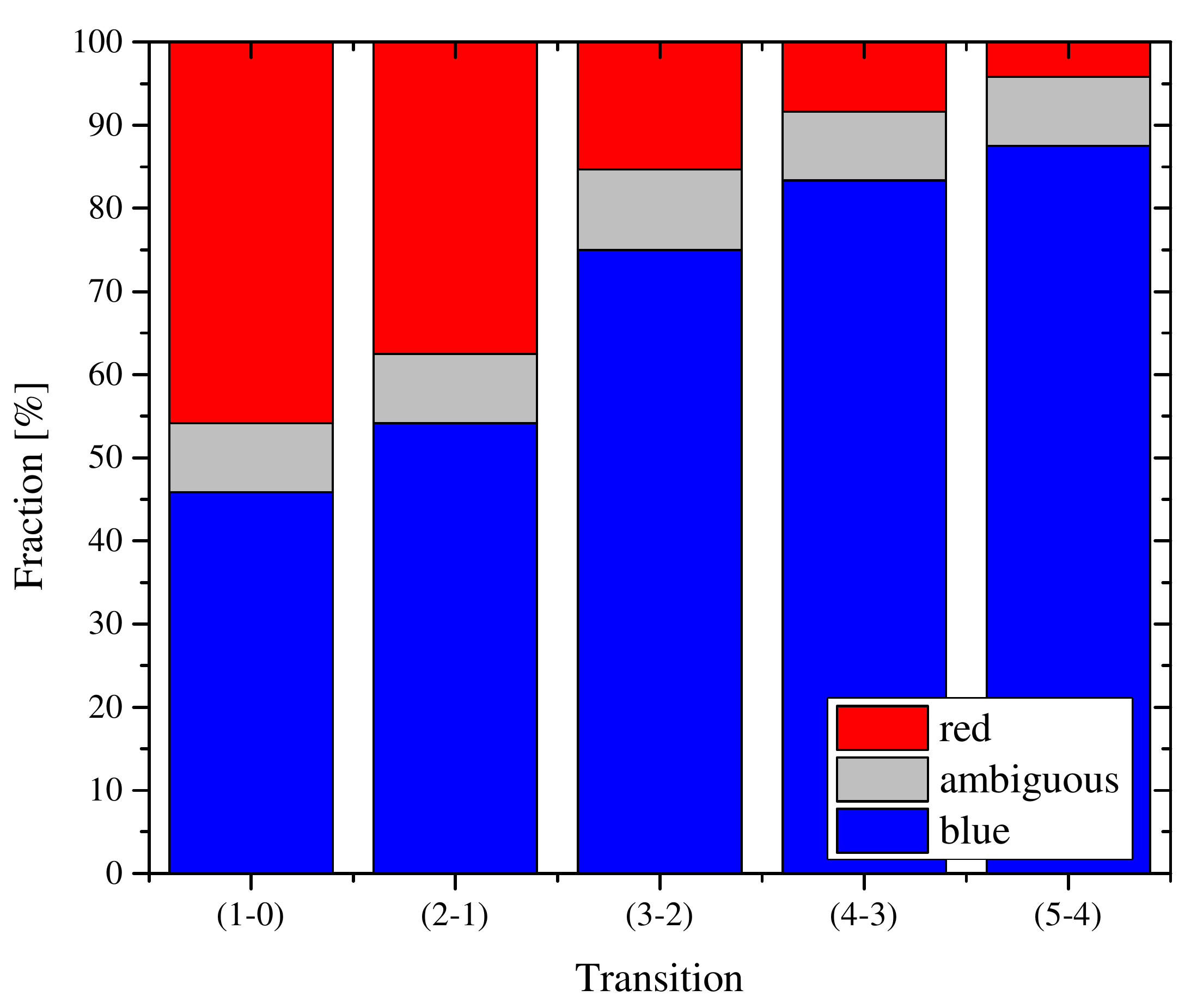}
			\caption{Results using the H$^{13}$CO$^+$ (4-3) transition as velocity reference.}
			\label{fig_04_deltav_h13co43}
		\end{subfigure} 
		\caption{Chart bars of summarising the results of $\delta v$ analysis. The graphs give the fractions of blue, red asymmetric or ambiguous line profiles at each transitions for all optically thick tracers, cores and sight-lines. The total number per transition is 72. The optically thin species and transitions which have been used as references are given in the corresponding captions. With increasing transition the numbers of blue line profiles increase, as the numbers of red asymmetric line profiles decreases. The fraction of ambiguous line profiles is almost constant. }
		\label{fig_04_deltav}
	\end{figure*}

	\begin{table*}
	\centering
	\caption{Summary of $\delta v$ analysis with different optically thin reference lines. The columns show the mean and median values of $\delta v$ and the fraction of blue, ambiguous and red asymmetric line profiles. The total number per transition is 36. The summary of these data is shown in Fig.\ \ref{fig_04_deltav}.}
	\begin{subtable}{0.48\textwidth}
		\caption{Results using the N$_2$H$^+$ (1-0) transition as velocity reference.}
		\label{tab02_deltav_n2h10}
		\begin{tabular}{lc|cc|ccc}
		\hline
		\multicolumn{2}{c|}{transition} & mean $\delta v$ & median $\delta v$ & blue & ambiguous & red \\ \hline 
		& & \multicolumn{2}{c|}{km s$^{-1}$} & \multicolumn{3}{c}{\%} \\ \hline \hline 
		HCN & (1-0) & 0.00821 &  -0.341 &    52.8 &    5.56 &    41.7 \\
		& (2-1) &  -0.298 &  -0.731 &    58.3 &    13.9 &    27.8 \\
		& (3-2) &  -0.718 &  -0.822 &    77.8 &    13.9 &    8.33 \\
		& (4-3) &  -0.625 &  -0.598 &    88.9 &    5.56 &    5.56 \\
		& (5-4) &  -0.586 &  -0.543 &    80.6 &    13.9 &    5.56 \\ \hline
		HCO$^+$ & (1-0) &  0.0504 &   0.163 &    38.9 &    11.1 &    50.0 \\
		& (2-1) & -0.0549 &    0.00 &    47.2 &    8.33 &    44.4 \\
		& (3-2) &  -0.483 &  -0.798 &    66.7 &    11.1 &    22.2 \\
		& (4-3) &  -0.499 &  -0.618 &    72.2 &    16.7 &    11.1 \\
		& (5-4) &  -0.656 &  -0.681 &    83.3 &    11.1 &    5.56 \\ \hline
		\end{tabular}
	\end{subtable} \hfill
	\begin{subtable}{0.48\textwidth}
		\caption{Results using the N$_2$H$^+$ (3-2) transition as velocity reference.}
		\label{tab02_deltav_n2h32}
		\begin{tabular}{lc|cc|ccc}
		\hline
		\multicolumn{2}{c|}{transition} & mean $\delta v$ & median $\delta v$ & blue & ambiguous & red \\ \hline 
		& & \multicolumn{2}{c|}{km s$^{-1}$} & \multicolumn{3}{c}{\%} \\ \hline \hline 
		HCN & (1-0) &   0.175 &  -0.115 &    52.8 &    5.56 &    41.7 \\
		& (2-1) &  -0.199 &  -0.766 &    58.3 &    13.9 &    27.8 \\
		& (3-2) &  -0.711 &  -0.917 &    80.6 &    11.1 &    8.33 \\
		& (4-3) &  -0.532 &  -0.565 &    88.9 &    2.78 &    8.33 \\
		& (5-4) &  -0.501 &  -0.539 &    83.3 &    5.56 &    11.1 \\ \hline
		HCO$^+$ & (1-0) &   0.304 &   0.304 &    38.9 &    11.1 &    50.0 \\
		& (2-1) &   0.168 &    0.00 &    47.2 &    8.33 &    44.4 \\
		& (3-2) &  -0.447 &  -0.697 &    66.7 &    11.1 &    22.2 \\
		& (4-3) &  -0.456 &  -0.610 &    72.2 &    13.9 &    13.9 \\
		& (5-4) &  -0.608 &  -0.665 &    83.3 &    8.33 &    8.33 \\ \hline
		\end{tabular}
	\end{subtable} \\ \medskip
	\begin{subtable}{0.48\textwidth}
		\caption{Results using the H$^{13}$CO$^+$ (3-2) transition as velocity reference.}
		\label{tab02_deltav_h13co32}
		\begin{tabular}{lc|cc|ccc}
		\hline
		\multicolumn{2}{c|}{transition} & mean $\delta v$ & median $\delta v$ & blue & ambiguous & red \\ \hline 
		& & \multicolumn{2}{c|}{km s$^{-1}$} & \multicolumn{3}{c}{\%} \\ \hline \hline 
		HCN & (1-0) & -0.0241 &  -0.298 &    52.8 &    5.56 &    41.7 \\
		& (2-1) &  -0.407 &  -0.982 &    58.3 &    13.9 &    27.8 \\
		& (3-2) &   -1.01 &   -1.24 &    83.3 &    8.33 &    8.33 \\
		& (4-3) &  -0.832 &  -0.841 &    91.7 &    2.78 &    5.56 \\
		& (5-4) &  -0.805 &  -0.907 &    88.9 &    8.33 &    2.78 \\ \hline
		HCO$^+$ & (1-0) &  0.0905 &   0.138 &    38.9 &    11.1 &    50.0 \\
		& (2-1) & -0.0373 &    0.00 &    47.2 &    8.33 &    44.4 \\
		& (3-2) &  -0.677 &  -0.998 &    66.7 &    11.1 &    22.2 \\
		& (4-3) &  -0.672 &   -1.04 &    75.0 &    13.9 &    11.1 \\
		& (5-4) &  -0.904 &  -0.990 &    86.1 &    8.33 &    5.56 \\ \hline
		\end{tabular}
	\end{subtable} \hfill
	\begin{subtable}{0.48\textwidth}
		\caption{Results using the H$^{13}$CO$^+$ (4-3) transition as velocity reference.}
		\label{tab02_deltav_h13co43}
		\begin{tabular}{lc|cc|ccc}
		\hline
		\multicolumn{2}{c|}{transition} & mean $\delta v$ & median $\delta v$ & blue & ambiguous & red \\ \hline 
		& & \multicolumn{2}{c|}{km s$^{-1}$} & \multicolumn{3}{c}{\%} \\ \hline \hline 
		HCN & (1-0) & -0.0371 &  -0.311 &    52.8 &    5.56 &    41.7 \\
		& (2-1) &  -0.473 &   -1.10 &    61.1 &    11.1 &    27.8 \\
		& (3-2) &   -1.04 &   -1.30 &    83.3 &    8.33 &    8.33 \\
		& (4-3) &  -0.850 &  -0.908 &    91.7 &    2.78 &    5.56 \\
		& (5-4) &  -0.829 &  -0.921 &    88.9 &    8.33 &    2.78 \\ \hline
		HCO$^+$ & (1-0) &  0.0889 &   0.137 &    38.9 &    11.1 &    50.0 \\
		& (2-1) & 0.00821 &    0.00 &    47.2 &    5.56 &    47.2 \\
		& (3-2) &  -0.696 &   -1.02 &    66.7 &    11.1 &    22.2 \\
		& (4-3) &  -0.682 &   -1.05 &    75.0 &    13.9 &    11.1 \\
		& (5-4) &  -0.929 &   -1.03 &    86.1 &    8.33 &    5.56 \\ \hline
		\end{tabular}
	\end{subtable}
	\end{table*}

\subsection{Line Profile Asymmetries}\label{res_deltav}
	In this section we quantify the characterisation of the line profiles of optically thick species. 
	For this purpose we employ the normalised velocity difference $\delta v$ discussed in Sect.\ \ref{meth_deltav}. 

	Following S12 we start our analysis by using N$_2$H$^+$ (1-0) as reference line.
	Table \ref{tab02_deltav_n2h10} lists the mean and median $\delta v$ as well as the fraction of line profiles in our sample classified as blue, red asymmetric and ambiguous for each transition of each optically thick species.
	Fig.\ \ref{fig_04_deltav_n2h10} summarises the results. 
	It shows the fraction of blue, red asymmetric and ambiguous line profiles at the individual transition.
	The numbers are summed up for all optically thick tracers.
	In both the table and figure, the majority of line profiles are not blue asymmetric, but either red asymmetric or ambiguous in the case of the (1-0) transitions. 
	However, the fraction of blue asymmetric line profiles increases towards higher transitions, and the fraction of red asymmetric line profiles decreases. 
	The number of ambiguous line profiles is approximately constant.
	The mean and median values of $\delta v$ tend to become more negative at higher transitions.

	Considering the whole sample we detect the maximal number of blue asymmetric spectra at the (5-4) transition (81.9\%; \linebreak in HCO$^+$ 83.3\%).
	In the case of HCN the greatest fraction of blue asymmetric line profiles (88.9\%) is at the (4-3) transition.
	Regardless of which transition we observe the cores, the line profiles are never solely blue asymmetric.

\subsubsection{Dependence on Reference Line}\label{res_optthin}
	Before we continue our discussion we test how $\delta v$ depends on the reference line.
	For this we repeat the analysis above using\linebreak N$_2$H$^+$ (3-2), H$^{13}$CO$^+$ (3-2) and H$^{13}$CO$^+$ (4-3).
	These lines have the advantages that they are still sufficiently bright to be clearly detected above the noise (see Sect.\ \ref{res_obs}) and are probably less optically thick than the (1-0) transitions (see Sect.\ \ref{res_tausurf}).

	The results are given in Tables \ref{tab02_deltav_n2h32}, \ref{tab02_deltav_h13co32} \& \ref{tab02_deltav_h13co43} and \linebreak Figs.\ \ref{fig_04_deltav_n2h32}, \ref{fig_04_deltav_h13co32} \& \ref{fig_04_deltav_h13co43}.
	Here we see the same trend as before.
	At lower transitions the fraction of total blue asymmetric line profiles is less than 50\%. 
	When going to higher transitions the fractions increase continuously up to 83.3\% (N$_2$H$^+$ (3-2)) and 87.5\% \linebreak (H$^{13}$CO$^+$ (3-2) \& (4-3)).
	Correspondingly, the mean and median values of derived $\delta v$ become more negative.
	The fraction of red asymmetric line profiles decreases as the transition increases.
	Again the number of ambiguous line profiles is approximately constant.
	Thus, we improve our reliability, even if only slightly, by using either N$_2$H$^+$ (3-2), H$^{13}$CO$^+$ (3-2) or H$^{13}$CO$^+$ (4-3) as references.
	
	Going to higher transitions increases the reliability of observing a blue asymmetric profile from a collapsing core not only in the optically thick tracers, but also in the optically thin reference lines.
	The best results are gained when using H$^{13}$CO$^+$ (3-2) and (4-3) as reference lines, even if the results here differ slightly from those derived with N$_2$H$^+$ (1-0) and (3-2) (and only marginally from each others).

\subsubsection{Remarks on the Influence of Abundance Models}\label{res_abun}

	\begin{figure*}
		\begin{minipage}{\textwidth}
			\centering
			\begin{subfigure}{0.49\textwidth}
				\centering
				\includegraphics[width=\textwidth]{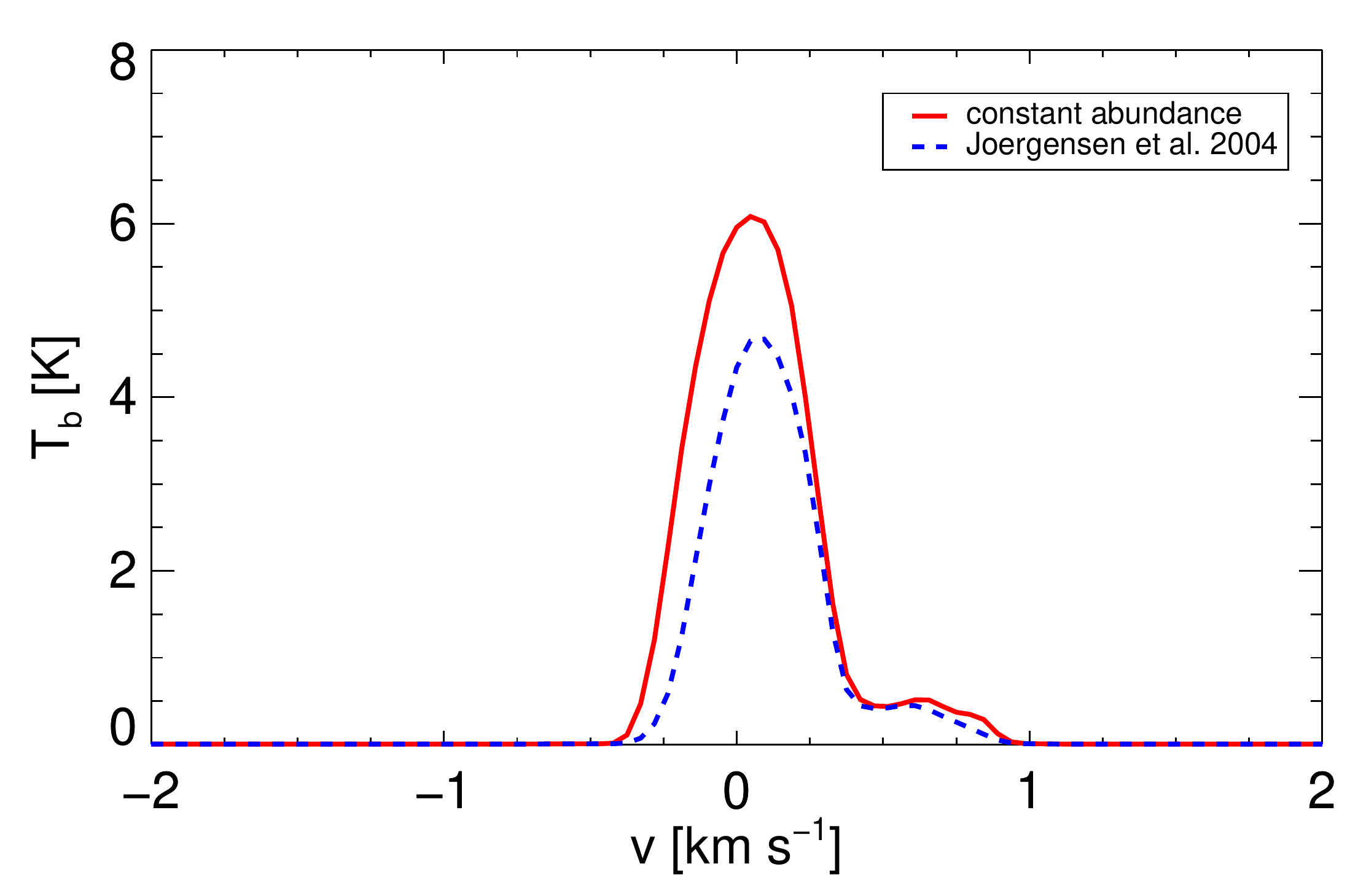}
				\caption{Core C at $i$ = 315$\degr$, $\phi$ = 0$\degr$}
				\label{fig_05_abun_non}
			\end{subfigure}
			\hfill			
			\begin{subfigure}{0.49\textwidth}
				\centering
				\includegraphics[width=\textwidth]{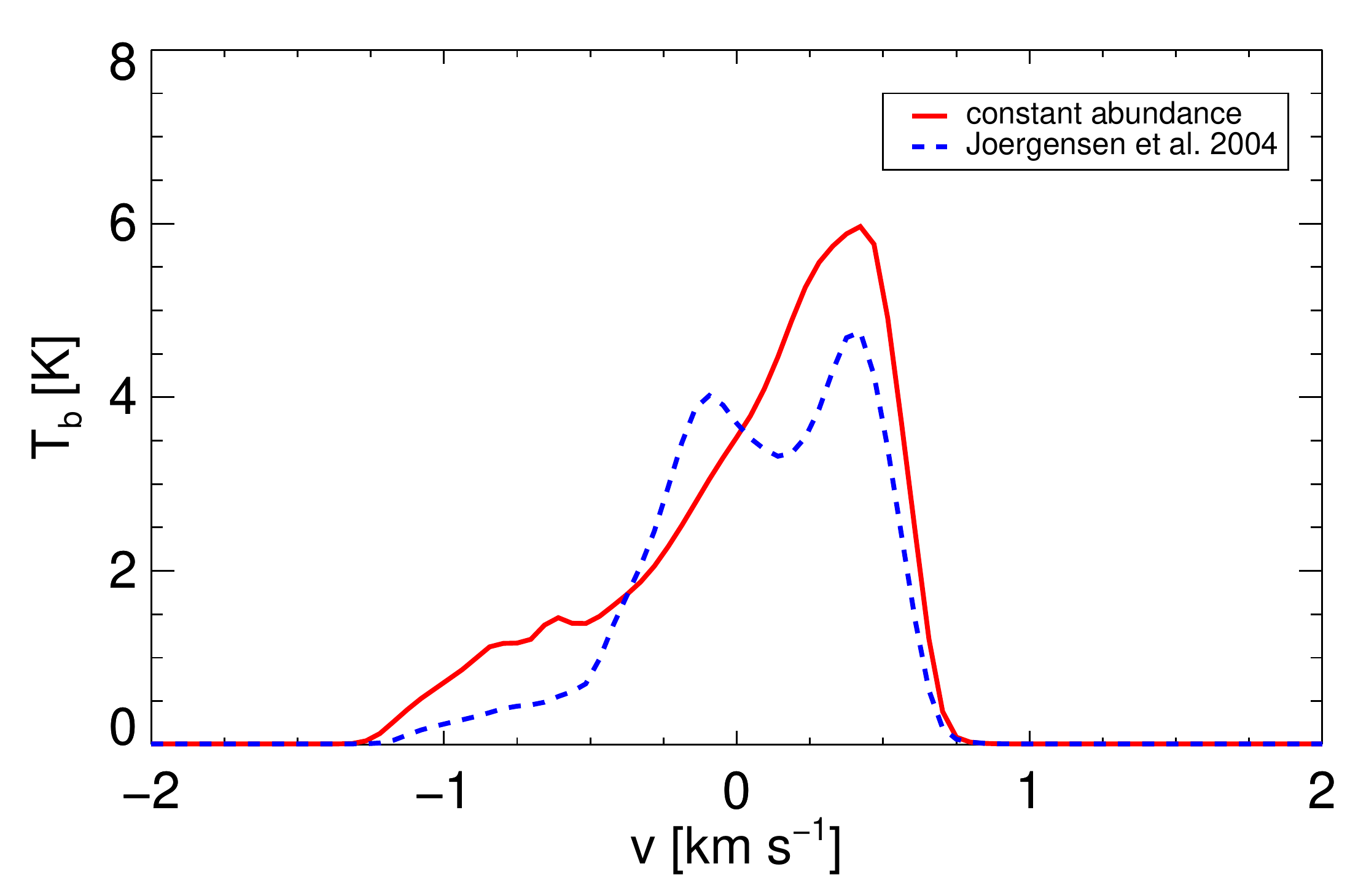}
				\caption{Core A at $i$ = 135$\degr$, $\phi$ = 0$\degr$}
				\label{fig_05_abun_change}
			\end{subfigure}
			\caption{Examples for non-/changing line profile asymmetries. 
			In both figures the red bold lines show the HCO$^+$ (3-2) line profiles that are modelled assuming HCO$^+$ to be constant abundant (c-abundant).
			The blue dashed lines represent the line profiles that result when the abundances of HCO$^+$ follows a step-function as described by \citet[J-abundant]{Joergensen2004}. 
			\textit{Left:} HCO$^+$ (3-2) profiles in Core C at $i$ = 315$\degr$, $\phi$ = 0$\degr$.
			The J-abundant line profile is less bright than the c-abundant line profile.
			Still, both have the same line profile asymmetry.
			\textit{Right:} HCO$^+$ (3-2) profiles in Core A at $i$ = 135$\degr$, $\phi$ = 0$\degr$.
			Here, the line profiles differ clearly from each other. 
			The c-abundant line profile has other a blue shoulder than a blue peak that cannot be resolved.
			The J-abundant line profile, however, is clearly double-peaked and red asymmetric. }
			\label{fig_05_abun}
		\end{minipage}
	\end{figure*}

	As mentioned in Sect. \ref{meth_tracer} we assume constant abundances, relative to H$_2$, for all of our molecular tracers.
	However, observations have shown that depletion occurs at some point for all molecules, especially in dense environments like we investigate here \citep[regions with n(H$_2$) > 10$^6$ cm$^{-3}$, see e.\ g.][]{Pratap1997,Bergin1997,Bergin2002,Friesen2010}.
	
	We test the influence of depletion on the HCO$^+$ results using the abundance models of \citet{Joergensen2004}. 
	In their model the abundances of carbon bearing molecules, like CO, C$^{18}$O and HCO$^+$, can be described by a step-function.
	The step-function distinguish between the warm-diffuse and cold-dense parts of the interstellar medium.
	Latter is defined as the gas where n(H$_2$) > 3 $\times$ 10$^4$ cm$^{-3}$ and \linebreak T(gas) < 30 K.
	The abundances within these regions are $\chi_0$ and $\chi_D$, respectively, whereas $\chi_0$ > $\chi_D$.
	Unfortunately, \citeauthor{Joergensen2004} offer only the values for C$^{18}$O.
	However, we can use the given relation between abundances of C$^{18}$O, CO and HCO$^+$ to compute the values for HCO$^+$.
	These are \linebreak $\chi_0$(HCO$^+$) = 5.45 $\times$ 10$^{-9}$ and $\chi_D$(HCO$^+$) = 5.45 $\times$ 10$^{-10}$.
	\newpage

	We see that the shapes and asymmetries of the individual line profiles may change in some cases.
	Fig.\ \ref{fig_05_abun} we illustrate this with two examples.
	The red bold lines show the line profiles that were constructed using a constant abundance (c-abundant line profiles, hereafter). 
	The line profiles following the step-like abundance as described by \citeauthor{Joergensen2004} (hereafter, J-abundant line profiles) are sketched with blue dashed lines.
	Fig.\ \ref{fig_05_abun_non} shows the line profiles in Core C at $i$ = 315$\degr$ and \linebreak $\phi$ = 0$\degr$. 
	The J-abundant line profile is less bright than the c-abundant line profile.
	This is due to the smaller amount of HCO$^+$ in the dense regions.
	Our J-abundant line profiles have, on average, a mean brightness temperature of 4.2 K, whereas the c-abundant line profiles have an average mean brightness temperature of \linebreak 4.6 K (see Table \ref{tab03_Tmb}).
	Nevertheless, both line profiles that we show here share the same shape.
	In Fig.\ \ref{fig_05_abun_change} the models differ from each others.
	Here we see the line profiles in Core A at \linebreak $i$ = 135$\degr$ and $\phi$ = 0$\degr$.
	The c-abundant line profile has a much more pronounced red peak and the blue peak transformed to a less marked blue shoulder.
	The J-abundant shows a well-defined doubled-peaked line profile that is slightly red asymmetric.
	Thus, if one is interested in reconstructing the exact spectra of a specific object, the effects of depletion need to be considered.

	However, the over-all statistics of HCO$^+$ line profiles in our sample do not significantly change with the abundance model. 
	The fraction of blue / red asymmetric / ambiguous line profiles at the (3-2) transition is now 69.5 / 8.3 / 22.2\%.
	That agrees with the previous fractions of (blue / red / ambiguous =) 66.7 / 11.1 / 22.2\% that we obtained with constant abundances of HCO$^+$ and \linebreak H$^{13}$CO$^+$ (3-2) as reference line (see Sect.\ \ref{res_optthin}).
	
	We conclude at this point that depletion is not neglectable in principle, but it does not change the general behaviour of infall signatures in complex pre-stellar cores.
	Since we are focusing on how the non-trivial geometry and dynamics of the gas may affect the line profiles we continue our investigations assuming constant abundances for all molecular tracers. 
	However it would be interesting in future works to consider both the effects of complex geometries and chemical inhomogeneities upon line profiles within the same simulation using a full chemical network with depletion.

\subsection{The Effect of Observational Noise}\label{res_obs}
	So far, we have ignored observational limitations in the analysis.
	Real observed line profiles are not only affected by optical depth effects along the line of sight, but also by noise and limited resolution.
	In this section we aim to investigate the influence of noise on the line profile asymmetry analysis.
	\newpage

	Noise is a constant problem in observational astronomy, especially when one wants to study weak lines.
	The measured noise levels depends on many different factors, like the instruments used, the conditions during observations, the brightness of the source etc.
	Here, we do not go into detail about possible noise sources, but examine the influences noise may have on our results.

	Hereafter we choose an average noise level in brightness temperatures of about $\sigma$ = 0.2 K that is reasonable for observations in star-forming regions \citep[e.\ g.,][]{Maezawa1999}.
	If we would like to observe our transition lines with this noise level and a resolution that suits our spectra here \linebreak ($\Delta v$ = 0.047 km s$^{-1}$) we would need exposure times between some minutes and an hour at the IRAM 30m telescope\footnotemark[2].
	\footnotetext[2]{\url{https://mrt-lx3.iram.es/nte/time_estimator.psp}}

	All the lines that we consider, except for some of \linebreak H$^{13}$CO$^+$ (5-4), would be detected above 5$\sigma$ noise.
	As \linebreak H$^{13}$CO$^+$ (3-2) \& (4-3) are not affected by this problem we still suggest them as the optimal reference lines.
	However, we see significant differences between minimal and maximal detected brightness temperatures in each transitions, since the amount of visible gas depends on the line of sight-angle at which one looks at the cloud.
	The brightness temperatures of the (1-0) transitions are between 1 and 14 K, whereas those of (5-4) are between 0.3 and 2.4 K (see Table \ref{tab03_Tmb} for mean brightness temperatures).

	For our classification of optically thick line profiles, however, we demand that the relative brightness temperature difference between both peaks is at least 10\%, see equation (\ref{def_amb_cr1a}).
	In absolute numbers these are temperature differences less than 1.4 K in the (1-0) transitions and less than 0.3 K in the (5-4) transitions.
	These small differences are very sensitive to noise disturbances. 
	There might be cases where noise affects one of the line profile components more than the other one such that the difference between the peaks changes.
	This effect would influence the classification of those line profiles, especially if the differences are already small as they are in ambiguous line profiles.

\begin{figure}
		\centering
		\includegraphics[width=0.49\textwidth]{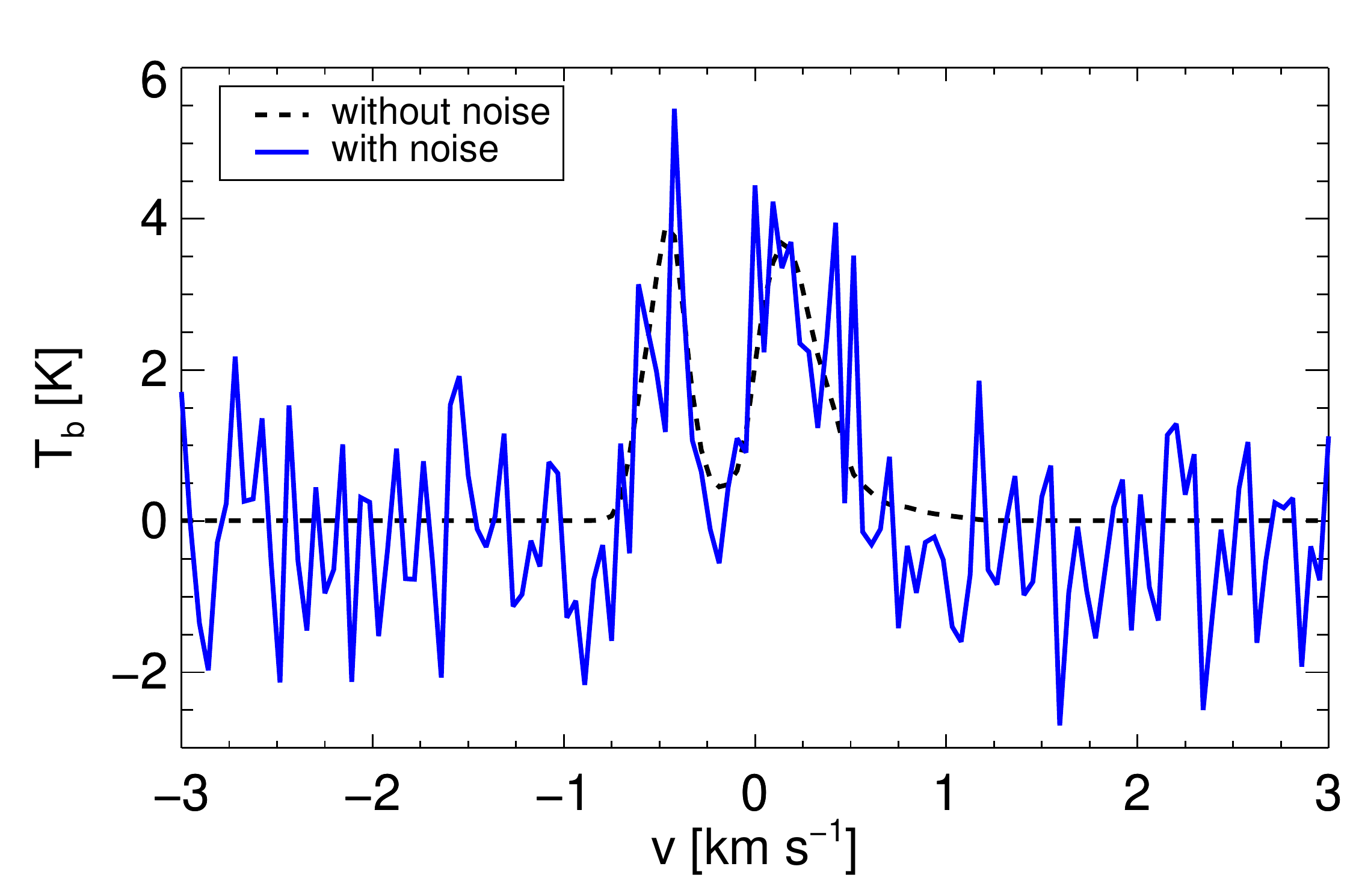}
		\caption{Example of the influence of noise on the classification of ambiguous line profiles. The dashed black line shows the an exemplary ambiguous line profile of the HCN (3-2) transition in Core A at $i$ = 315$\degr$ and $\phi$ = 0$\degr$ without noise. The blue line shows the same line profile with noise which has been generated by adding a random, normalised noise distribution to the HCN line profile. The noisy line profile seems to be blue asymmetric.}
		\label{fig_06_noise}
	\end{figure}

	In Fig.\ \ref{fig_06_noise} we illustrate this by using the example of the ambiguous line profile of HCN (3-2) in Core A at $i$ = 315$\degr$ and \linebreak $\phi$ = 0$\degr$ (dashed black line).
	We simulate noise by using a normalised random distribution that we average to the typical noise level in star-forming regions and add it to the HCN line profile (blue line).
	The difference between the peak brightness temperatures increases and becomes greater than 10\%.
	We would classify the line profile as asymmetric, although it is ambiguous.
	
	Of course, this works also the other way around as noise reduces the differences between the peaks in asymmetric line profiles such that we would classify them as ambiguous.
	Thus, it is necessary to rephrase the peak brightness temperature differences related criterion, that is represented by equation (\ref{def_amb_cr1a}), so that it refers to the noise level of the observed region.
	The goal is to introduce a criterion which can be applied in observational studies and which returns reliable results that are not (significantly) distorted by noise.
	We again repeat the analysis above (see Sect.\ \ref{res_deltav}), but use the following criteria for ambiguous profiles:
	\begin{align}
		|T_{b,blue} - T_{b,red}|\,<\,n\,\cdot\,\sigma \label{def_amb_cr1b} \\ \vspace*{0.5\baselineskip}
		| v_{thick,blue} - v_{thick,red} | < \Delta v_{thin} \label{def_amb_cr2b}
	\end{align}\label{def_amb_noise}
	\noindent where $n$ is an integer that symbolises the degree of significance above the noise level.

	We investigate levels of 3$\sigma$, 5$\sigma$ and 6$\sigma$.
	Fig.\ \ref{fig_06_noise} shows a summary chart of the results obtained with 5$\sigma$.
	The numbers are summed up for all optically thick tracers and optically thin reference lines such that the plotted fractions only depend on the transition of optically thick species.
	Although we only offer the results which are derived with 5$\sigma$, the results using the 3$\sigma$ and 6$\sigma$ thresholds are similar, as discussed below.

	\begin{figure}
		\centering
		\includegraphics[width=0.49\textwidth]{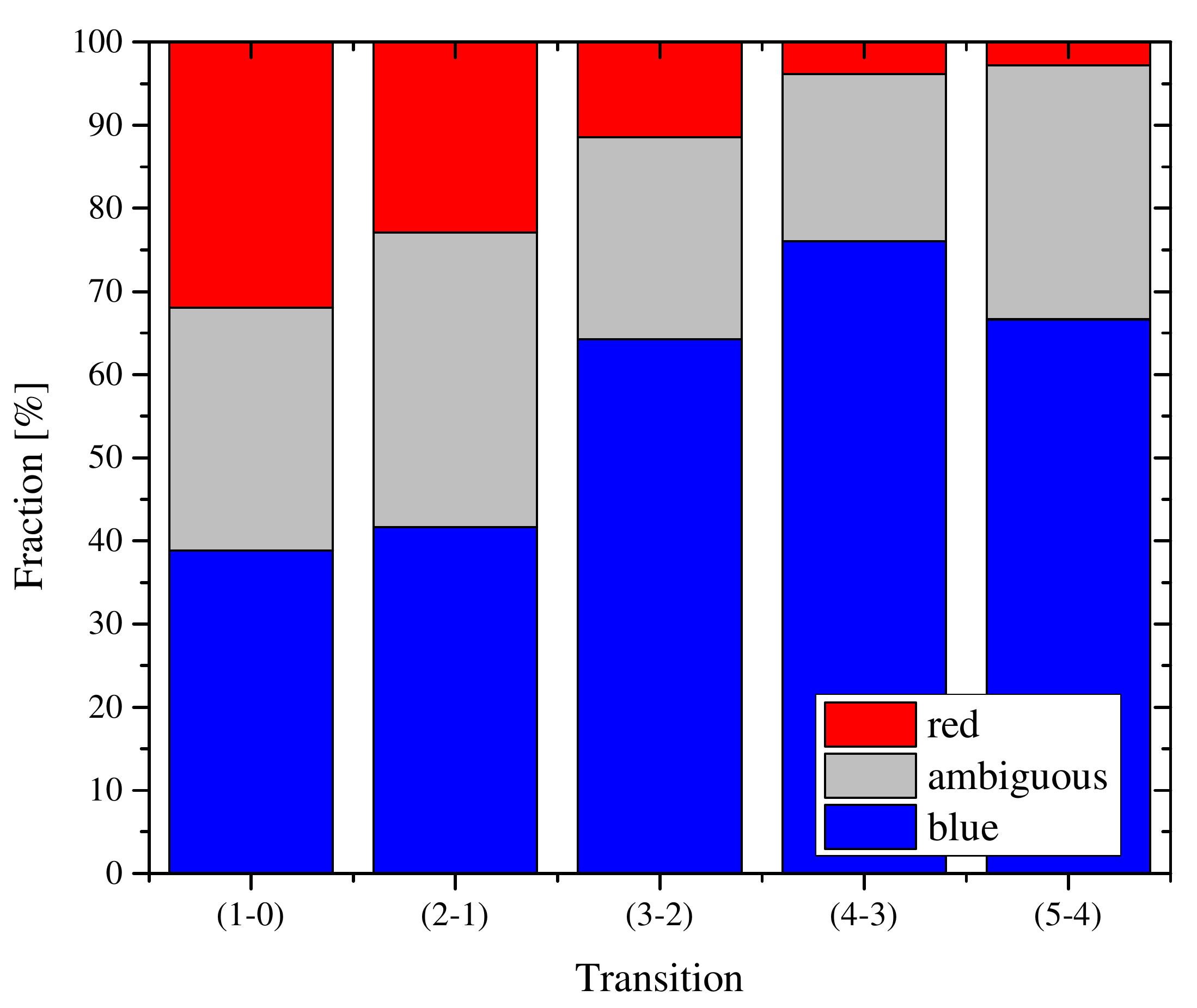}
		\caption{Chart bar summarising the results of $\delta v$ analysis using a 5$\sigma$ noise level as threshold. Analogously as in Fig.\ \ref{fig_04_deltav}, but this chart bar sums up the results obtained with all optically thick molecular tracer, as well as optically thin reference lines. The total number per transition is 288. The chart bar shows again that the fraction of blue line profiles increases with increasing transition, as the numbers of red profiles decrease. The fraction of ambiguous line profiles has become large compared to the study without noise criterion.}
		\label{fig_07_deltav_sigma5}
	\end{figure}

	The figure shows the same trends as the previous results.
	In the (1-0) transitions the fraction of blue asymmetric spectra is on the order of 40\% which increases up to 60 -- 75\% in the (5-4) transitions.
	These numbers are about 5 -- 10\% lower than the numbers derived in Sect.\ \ref{res_deltav}.
	Similarly, the total fraction of red asymmetric spectra decreases from about 40\% in the (1-0) transitions down to \linebreak 3 -- 5\% which is far less when in the investigations before.
	Consequently the total number of ambiguous line profiles increases dramatically by about 10\% compared to the results before.

	We can easily understand this if we consider the first criterion for ambiguous line profiles again. 
	When we have used a relative temperature difference limit of 10\%, see equation (\ref{def_amb_cr1a}), the threshold for detected asymmetries has been smaller than or on the same order of magnitude as the typical noise level in the case of weak lines. 
	By introducing a noise dependent formulation of this criterion, see equation (\ref{def_amb_cr1b}), we increase the threshold in most of the cases and select only those line profiles with higher peak temperature differences and, thus, more obvious line profile asymmetries.
	Furthermore, we also classify the other spectra with peak differences between 10\% of the higher emission peak and 3/5/6$\sigma$ as ambiguous now.
	
	Hence, it is intuitive that the fraction of ambiguous spectra becomes larger with higher $n$.
	In all transitions the numbers of ambiguous line profiles doubles when we use the 6$\sigma$ threshold compared to the numbers derived with the 3$\sigma$ limit.
	At the same time the numbers of blue and red asymmetric line profiles decreases with higher $n$, though not as rapidly as the numbers of ambiguous spectra increase.

	In summary, rephrasing our first criterion for ambiguous spectra such that the minimal difference of peak brightness temperatures depends on an absolute scale, like the noise level, has an impact on the results of $\delta v$ analysis.
	The fraction of ambiguous spectra becomes larger, since the threshold for asymmetric line profiles is higher now and fainter lines are, therefore, preferentially classified as ambiguous.
	The results seem to be worse than those presented in Sect.\ \ref{res_deltav}, where we used a more idealised ansatz.
	But a lower number of red asymmetric line profiles also results in a lower number of "false" detections which improves the general outcome of the analysis.

	\begin{figure*}
		\begin{minipage}{\textwidth}
			\centering
			\includegraphics[width=0.95\textwidth]{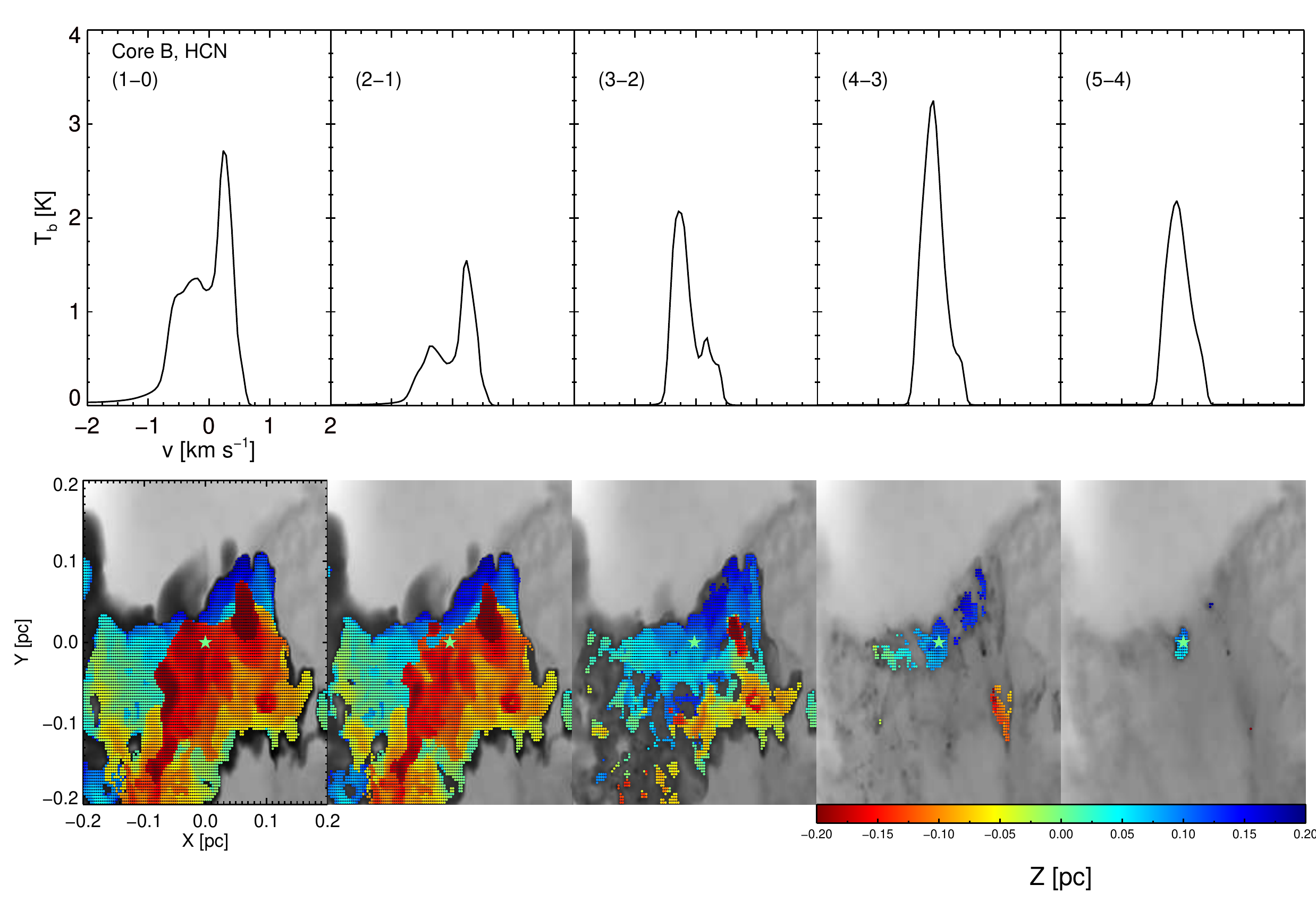}
			\caption{Line profiles and optical depth surfaces of HCN in Core B at $i$ = 90$\degr$ and $\phi$ = 135$\degr$. 
			\textit{Upper panel:} line profiles of HCN (1-0) to (5-4). 
			\textit{Lower panel:} corresponding optical depth surfaces (see Sect.\ \ref{meth_tausurf}). 
			The grey-scale background shows the emission of HCN at the different transitions, respectively. 
			The colour scale in the foreground refers to the first points along the line-of-sight where the tracer becomes optically thick for the first time. 
			The positions are closer to the observer the redder the colour scale is.
			The green star points to the centre of the core.
			The areas of optically thick HCN becomes smaller at higher transitions (due to higher densities needed to populate these energy levels) and more closer to the core. 
			At the same time the line asymmetry becomes bluer until the red peak transforms into a shoulder to the blue peak.}
			\label{fig_08_multi_hcn}
		\end{minipage}
	\end{figure*}

	\begin{figure*}
		\begin{minipage}{\textwidth}
			\centering
			\begin{subfigure}{0.49\textwidth}
				\centering
				\includegraphics[width=\textwidth]{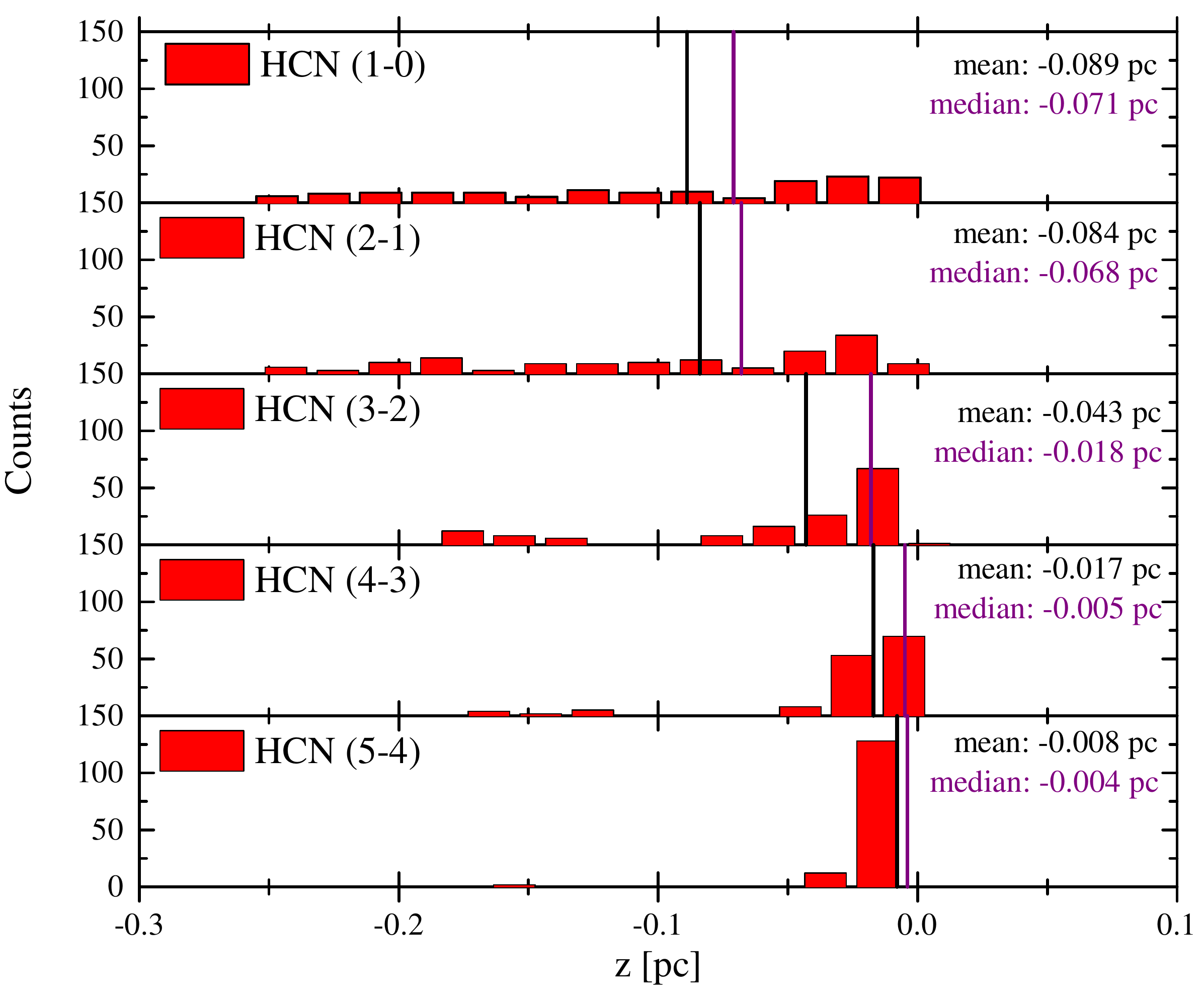}
				\caption{HCN}
				\label{fig_09_tausurf_histo_hcn}
			\end{subfigure}
			\hfill
			\begin{subfigure}{0.49\textwidth}
				\centering
				\includegraphics[width=\textwidth]{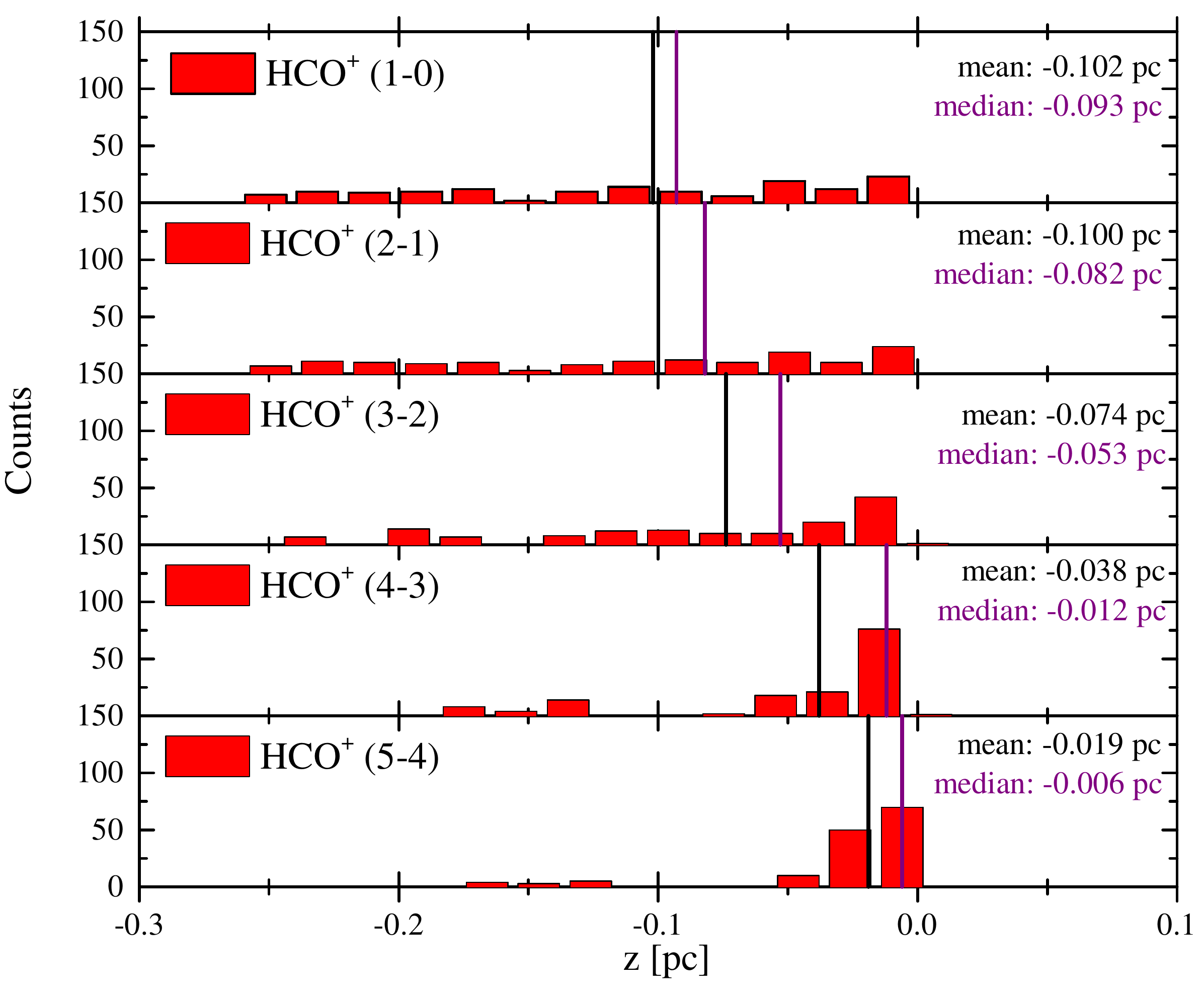}
				\caption{HCO$^+$}
				\label{fig_09_tausurf_histo_hco}
			\end{subfigure}
			\caption{Histograms of results of optical depth surface analysis of (\textit{left}:) HCN (\textit{right}:) and HCO$^+$. The histograms show the positions where the emission transitions becomes optically thick along the line of sight $z$. Negative $z$ are in the foreground of core according to the observer, positive $z$ in the background. The core centre is at $z$ = 0. The numbers are summed up for all cores. Average (black) and median (violet) positions are also given. The histograms show that the regions with optically thick gas are distributed closer to the core at higher transitions, and the distributions become narrower.}
			\label{fig_09_tausurf_histo}
		\end{minipage}
	\end{figure*}

	\begin{figure*}
		\begin{minipage}{\textwidth}
			\centering
			\includegraphics[width=0.95\textwidth]{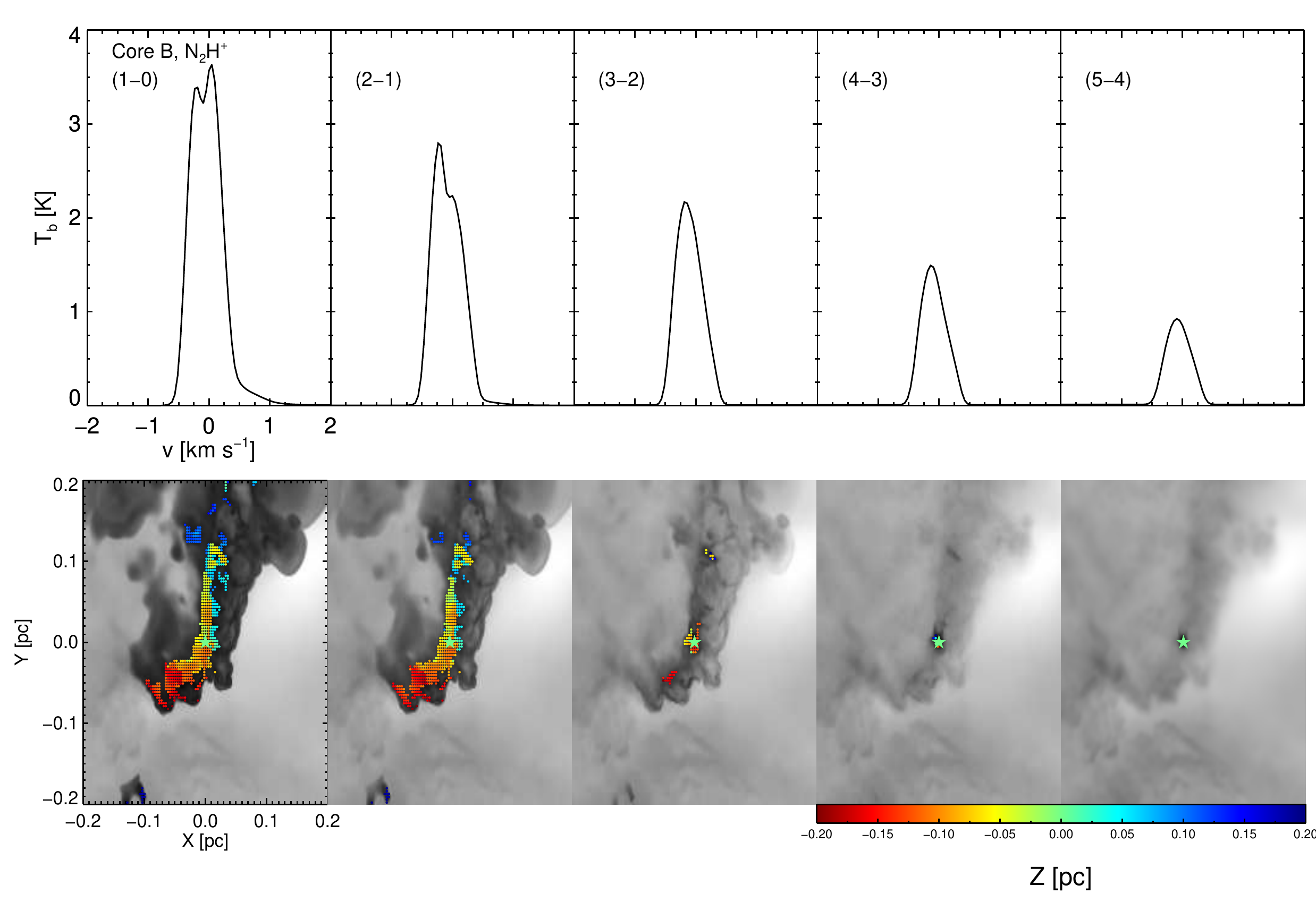}
			\caption{Line profiles and optical depth surfaces of N$_2$H$^+$ in Core B at $i$ = 0$\degr$ and $\phi$ = 0$\degr$. 
			\textit{Upper panel:} line profiles of N$_2$H$^+$ (1-0) to (5-4). 
			\textit{Lower panel:} corresponding optical depth surfaces (as in Fig.\ \ref{fig_08_multi_hcn}, see Sect.\ \ref{meth_tausurf}).
			At low transitions N$_2$H$^+$ is optically thick and its line profiles are affected by self-absorption. 
			At higher transitions N$_2$H$^+$ thins out and the line profiles become more Gaussian shaped.}
			\label{fig_10_multi_n2h}
		\end{minipage}
	\end{figure*}

	\begin{table}
	\centering
	\caption[Summary of measured brightness temperatures.]{Summary of measured brightness temperatures. The table gives the average values of brightness temperatures. In the case of optically thick tracers the given values of brightness temperature correspond to the higher peak in asymmetric line profiles. If the line profiles are classified as ambiguous, both peak brightness temperatures were considered.}
	\begin{tabular}{c|cccc}
		\hline
		\backslashbox{Transition}{Species} & HCN & HCO$^+$ & N$_2$H$^+$ & H$^{13}$CO$^+$ \\ \hline
		 & \multicolumn{4}{c}{mean T$_{b}$ [K]} \\ \hline
		(1-0) & 4.3 & 6.1 & 4.1 & 2.1 \\
		(2-1) & 3.3 & 4.8 & 2.5 & 2.0 \\
		(3-2) & 3.1 & 4.6 & 2.5 & 1.5 \\
		(4-3) & 3.0 & 3.9 & 1.8 & 0.8 \\
		(5-4) & 2.3 & 3.4 & 1.0 & 0.3 \\ \hline
	\end{tabular}
	\label{tab03_Tmb}
	\end{table}

\subsubsection{Comparison to Observations}

	The line profiles of optically thick tracers have been widely examined in observational studies.
	\citet{Gregersen1997} observed 23 class 0 cores in HCO$^+$ (3-2).
	They found blue asymmetric line profiles in nine cores (39.1\%) and three cores (13.0\%) with red asymmetric line profiles.
	In a later study, \citet{Gregersen2000} investigated a sample of 17 starless cores.
	Half of these cores are either not detected in HCO$^+$ (3-2) or show optically thin line profiles.
	Only six (35.3\%) of the line profiles were blue asymmetric.
	If we would neglect the non-detections of \citeauthor{Gregersen2000}, the observed fraction of blue line profiles agrees with our results (around 64\% blue asymmetric (3-2) line profiles when using a 5$\sigma$ noise threshold, see Fig.\ \ref{fig_07_deltav_sigma5}).
	
	\citet{Andre2007} achieved a better detection rate.
	They observed 25 starless condensations with different optically thick and thin tracers, i.\ a.\ HCO$^+$ (3-2) and H$^{13}$CO$^+$ (1-0).
	\citeauthor{Andre2007} find 64\% of line profiles have blue asymmetries.
	The remaining 9 cores do not show blue infall signatures.
	This detection fraction is very similar to our predicted fraction (67\%) for HCO$^+$ (3-2) if all the cores are collapsing.

	\citet{Sohn2007} have observed a sample of cores in \linebreak HCN (1-0) and N$_2$H$^+$ (1-0).
	They selected sources from the sample of \citet{Lee1999} that are nearby (distances $\sim$ 100 pc), dense (n(H$_2$) $\sim$ 10$^{4-5}$ cm$^{-3}$), compact (r $\sim$ 0.05--0.35 pc) and have narrow N$_2$H$^+$ line widths ($\Delta v$(N$_2$H$^+$) $\sim$ 0.2--0.4 km s$^{-1}$).
	These properties are akin to the characteristics of our regions.
	In their analysis \citeauthor{Sohn2007} discuss the asymmetries of the three HCN hyperfine lines.
	28 (43.8\%) out of 64 line profiles have at least one blue asymmetric hyperfine component, whereas the other are either ambiguous or optically thin.
	Analogously, 21.9\% (14/64) contain at least one red asymmetric hyperfine component.
	These numbers are in agreement with our results.
	If using the 5$\sigma$ threshold we see that about 40\% of the (1-0) line profiles are blue asymmetric \linebreak (see Fig.\ \ref{fig_07_deltav_sigma5}).
	The 30\% red asymmetric line profiles in our sample appear higher than the 21.9\% found by \citet{Sohn2007}. 
	However, 32.8\% of \citeauthor{Sohn2007}'s sample include spectra with mixed asymmetries in the different hyperfine components. 
	Since we exclude all hyperfine components but the HCN (J = 1-0) F(2-1) in our study it is likely that  this may be the origin of the discrepancy.
	We expect that that will mainly influence the number of red asymmetric line profiles, since the F(2-1) is the optically thickest hyperfine component and rather traces the gas motion in regions that are further out and more problematic.

	Our results agree well with the findings of observational studies.
	Most of the above studies investigated isolated cores and, thus, ignored the effects of filamentary structures along the line of sight on the observed line profiles. 
	As a consequence, they may have underestimated the number of collapsing cores.
	\textit{Herschel} has emphasised the presence of filaments in star-forming regions and shown that the number of isolated cores is very small. 
	Here we show that surrounding filaments have a huge impact on the shapes of line profiles. 
	Upcoming ALMA observations will have much better possibilities of resolving the geometry of collapsing cores and the environments they are embedded within.
	This will help to test the reliability of measured line profile asymmetries.

\subsection{Origin of Emission Features}\label{res_tausurf}
	We have already seen that the line profiles do not necessarily meet our expectations.
	Either the line profiles do not have the `right' asymmetry, i.\ e., the HCN (1-0) line in Fig.\ \ref{fig_03_red10_blue32}, or there are additional components in the line profiles, i.\ e., Core C in Fig.\ \ref{fig_02_multi_all_10}.
	We investigate the distributions of density, velocity and optical depth to study this phenomena in more detail.
	At first we study the origin of additional line profile components. 
	Using the density and velocity data obtained by the GMC simulation we find them in the physical structure of the clusters. 
	Due to the complex velocity pattern, which is described by S12, the emission of particular regions are Doppler shifted. 
	As a consequence, these shifted component form their own line beside the main transition line.

	To find the origin of non-blue line profiles asymmetries and the reason for the observed reversal we use the optical depth surfaces, ODS (see Sect.\ \ref{meth_tausurf}).
	The advantage of ODS is that they connect theoretical three-dimensional density and velocity data with observable maps. 
	Furthermore, looking at the $\tau$ = 1 surface allows us to distinguish between regions that do or do not contribute to the observed intensities.
	With this we mean that parts of the inner, denser regions also emit photons.
	But we must not necessarily observe those photons, since optically thick gas may surround these parts, and block or scatter photons out of the line of sight.
	Knowing where the border between optically thin and thick gas is helps finding the origin of line profile features and anomalies.

	As has already been mentioned in Sect.\ \ref{res_profiles}, one of the most common cases concerning line profile asymmetries of optically thick molecular tracers is that the line profiles are red asymmetric at lower transitions, but become blue asymmetric at higher transitions.
	Fig.\ \ref{fig_08_multi_hcn} illustrates this reversal using the example of HCN in Core B at $i$ = 90$\degr$ and $\phi$ = 135$\degr$. 
	Although we will only give this example, we see that both HCN and HCO$^+$ behave that way at various line of sight angles in all cores, provided that they undergo a reversal.
	The figure shows the line profiles of \linebreak HCN (1-0) -- (5-4) transition and the corresponding ODS. 
	We see that the line profiles of HCN (1-0) and (2-1) are red asymmetric, whereas the line profiles at higher transitions are blue asymmetric (or in the case of (5-4) approximately Gaussian due to less optically thick gas).
	In the ODS we observe that in the case of HCN (1-0) and (2-1) the majority of  gas becomes optically thick far in front of the core, but close behind the central core region at the higher transitions. 
	Furthermore, at the lower transitions there is a lot of gas which becomes optically thick although it is spatially too far away to be part of the collapsing central region around the core (which we define as the volume within a radius of 0.05 pc around the core centre for simplicity).
	This gas belongs to the filaments around the core. 

	At higher transitions the amount of optically thick gas shrinks and occurs closer to the inner part of the core.
	This is more obvious in Fig.\ \ref{fig_09_tausurf_histo} which shows the ODS for all transitions of HCN and HCO$^+$ as histograms observed within the beam \linebreak (FWHM = 0.01 pc) along the line of sight radius $z$.
	It also provides the mean and median positions where $\tau$ = 1 is reached for the corresponding species and transition.
	At low transitions the optically thick gas is distributed on a large area around the core.
	When increasing transitions the optically thick gas concentrates within the central region which is represented by the shift of mean and median positions forward the core centre.

	Thus, red line profiles can be associated with optically thick gas within filamentary structure in front of the actual core region.
	Since lower transitions have lower critical densities, filaments may become dense enough to populate these levels, emit and may even become optically thick. 
	To populate higher transitions, the gas needs to reach higher densities which are only achieved in dense cores, but not in filaments.
	Blue asymmetric line profile observed at higher transitions observe infall motion within the core, even if the core is irregularly shaped.
	
	However, one has to be cautious when increasing the transition due to an observational aspect.
	At higher transitions the tracer becomes optically thinner and the weaker peak of the line profile may transform into a shoulder of the main peak.
	Fig.\ \ref{fig_08_multi_hcn} illustrates how the red peak of the HCN (3-2) transition line turns into a red shoulder at higher transitions.
	Although going to higher transitions may help to minimise the disturbing influences of surrounding filaments on the line profile asymmetries, too high a transition makes it hard to observationally detect such features.
	However, this behaviour is uncommon for the (3-2) transition which is why we still recommend it for observing collapse motions.
	\smallskip

	This method also shows why the results obtained with the \linebreak (3-2) and (4-3) transitions of the optically thin species as reference lines are better than whose obtained with N$_2$H$^+$ (1-0).
	Since we expected our optically thin tracers to be optically thin all over the core region and filaments, the corresponding ODS should have been empty.
	But, as mentioned in Sect.\ \ref{res_profiles}, we see absorption feature in the spectra of these species at lower transition \citep[as they are also observed, e.\ g.,\ by][]{Friesen2010}. 
	Fig.\ \ref{fig_10_multi_n2h} illustrates this with the example of N$_2$H$^+$ transitions in \linebreak Core B at $i$ = 0$\degr$ and $\phi$ = 0$\degr$.
	The figure shows the spectra and ODS of the five studied transitions.
	Self-absorption clearly affects the line profiles at lower transitions; the (1-0) line is red asymmetric and the (2-1) blue asymmetric.
	This is due to the large amount of optically thick N$_2$H$^+$ at the front edge of the core region marked in the ODS. 
	Of course, there is less optically thick material and it is located closer to the core compared to the optically thick tracers. 
	But as we need N$_2$H$^+$ and H$^{13}$CO$^+$ to be optically thin to obtain the central velocity $v_{thin}$ for $\delta v$ calculation, such `abnormal' features in the line profiles worsen the fitting procedures and cause greater errors.
	At higher transitions the lines become Gaussian shaped and the ODS become empty.

	At this point we have to emphasise that we have excluded the influences of hyperfine structures in this study.
	The computed optical depths and brightness temperatures of our line profiles are artificially higher than they would be in nature. 
	However, this simplification is valid in the context of the current investigation, since our species have either hyperfine components that are sufficiently isolated from each other to not interfere \citep[even at higher transitions, e.\ g.,][]{Daniel2006,Daniel2007}, or are too close together to be observationally resolvable \citep[e.\ g.,][]{Loughnane2012}. 
	Especially if one considers the complex density and velocity pattern in our regions, the lines are more likely affected by emission from other locations along the line of sight \linebreak (see Sect.\ \ref{res_profiles}) than they would be by line confusion alone.
	Therefore, we do not expect any qualitative changes in the general behaviour of the asymmetry when hyperfine structures are included.

\section{Summary and Conclusions}\label{sec_conclusion}
	The goal of this study is to identify an optimal, reliable tracer for collapsing cores in dense, filamentary clusters.
	For this, we continue the analysis of \citet[S12]{Smith2012a} and investigate the question of whether we can expect blue infall line profiles in irregularly shaped, collapsing cores which are embedded in filaments.

	S12 select three cores out of a Giant Molecular Cloud simulation, because they are irregular, deeply embedded in filaments, already collapsing, and about to form stars in the near future. 
	This creates a sample of real star-forming regions with complex density and velocity patterns.
	However, the analysis of S12 is limited to only one optically thick and one optically thin transition. 
	We address this limitation by modelling the (1-0) -- (5-4) transitions of two optically thick (HCN, HCO$^+$) and two optically thin (N$_2$H$^+$, H$^{13}$CO$^+$) species from twelve different line of sights.
	We assume constant abundances and ignore depletion effects.
	We use the normalised velocity difference, $\delta v$, as defined in equation (\ref{equ_def_deltav}) to classify the line profile asymmetries of optically thick tracers.
	We also introduce criteria for ambiguous spectra, see equations (\ref{def_amb_cr1a}) and (\ref{def_amb_cr2a}), that sort out the cases without clearly recognisable asymmetries.
	We study the number density, velocity and optical thickness distribution to localise the origin of line profiles features.

	We find that we can expect blue asymmetric line profiles in irregular cores, as we can in spherical cores. 
	Although less than 50\% of simulated line profiles are blue asymmetric in the (1-0) transitions, the fraction of blue line profiles increase when going to higher transitions of the optically thick tracers.
	We obtain our best results in the (4-3) transitions of HCN and HCO$^+$ where about 90\% of the line profiles are blue asymmetric.
	The success rates of the (3-2) and (5-4) transitions are only marginally lower, but they contain physical or observational disadvantages.
	The (3-2) transition is more affected by the surrounding filaments, whereas the (5-4) transition is fainter (see Table \ref{tab03_Tmb}) and less optically thick.
	Thus, the (4-3) transition offers the best combination of detectability of blue line profiles and visibility above typical noise levels.
	This makes it the most reliable tracer for dense core gas motions in star-forming clusters.

	We confirm this result by introducing classification criteria which take typical noise levels of such regions into account.
	They are described by the equations (\ref{def_amb_cr1b}) and (\ref{def_amb_cr2b}).
	This method reduce the maximal fraction of blue asymmetric line profiles to 60\% -- 75\% (depending on the used noise threshold). 
	These results are in good agreement with observational studies of, for example, \citet{Gregersen1997}, \citet{Gregersen2000} and \citet{Sohn2007}.
	Still, the (4-3) transitions produce the most reliable results.

	We find a small but significant number of spectra that show subtle shoulder-like features in the higher (4-3) and (5-4) transitions as opposed to a clear double-peaked line profile.
	In some cases such features may require prohibitively high signal-to-noise to be detected in real observing conditions.
	We suggest that the brighter (3-2) transitions may be a better avenue in such conditions. 
	However, we urge caution in interpreting the (3-2) line profile, as discussed above.

	Furthermore, we investigate the origin of `false' line profile asymmetries by using the ODS. 
	These are localised in the filaments around the cores where the gas is dense enough to become optically thick and hide the core behind.
	Since higher transitions arise from denser material they become more optically thin within the filaments.
	Thus, the cores and their blue line profiles become detectable. 
	In addition, the ODS show that the optically thin tracers may become optically thick in the lower transitions. 
	Since they are needed as reference lines in the $\delta v$ analysis we also use higher transitions of the optically thin tracers to improve the analysis.
	The (3-2) transitions turn out to be the best choice, since they are usually optically thin, bright and have central frequencies close to the (4-3) transitions of the optically thick tracers.

	We briefly explore the effect of depletion upon the line profiles and find that depletion in dense gas may change the details of individual spectra. 
	However, the overall percentage of blue, red and ambiguous line profiles is largely unchanged.
	In this study we focus on the line profiles of collapsing clustered cores resulting purely from the complex gas dynamics and geometries within these regions. 
	This allows us to isolate the effects of geometry upon line profiles in a relatively simple scenario that can be more easily understood. 
	However, in future work it would be beneficial to combine the effects of geometry and dynamics with a more detailed model of chemical inhomogeneities within the dense cores.
	\smallskip
	
	In summary, we can expect to detect blue infall line profiles in irregular, collapsing cores.
	Dense filaments around the cores along the line of sight cause non-blue asymmetric line profiles of optically thick tracers and non-Gaussian line profiles of optically thin tracers.
	Using higher transitions can avoid these effects.
	We obtain the best results with the (4-3) transitions of the optically thick tracers while using the (3-2) transition of optically thin tracers as reference lines.
	Future observations with, e.\ g., ALMA will be able to resolve the shapes of line profile, including self-absorption signatures and shoulder-like features, more precisely.

\section*{Acknowledgments}
	In particular the authors thank Cornelis Dullemond for offering his radiative transfer code \texttt{RADMC-3D}, on which this work bases, and helpful discussions. 
	R.S.K. acknowledges support from the European Research Council under the European Community’s Seventh Framework Programme (FP7/2007-2013) via the ERC Advanced Grant {\em STARLIGHT} (project number 339177). 
	The research leading to these results have also been supported by the German Science Foundation (DFG) via the Collaborative Research Center SFB 881 {\em The Milky Way System} (subprojects B1, B2, and B5) as well as via the Priority Program SPP 1573 {\em Physics of the Interstellar Medium} via project SM 321/1-1.
	Furthermore, the authors say thank you to Paul C.\ Clark and Simon C.\ O.\ Glover for many stimulating scientific discussions.

	\bibliographystyle{mn2e}
	\bibliography{ref}

\label{lastpage}
\end{document}